\documentclass[12pt]{article}
\usepackage{epsfig}
\usepackage{color}
\usepackage{amssymb,amsmath}
\usepackage{graphicx}
\usepackage{epsfig}

\newcommand{\bea}{\begin{eqnarray}}
\newcommand{\eea}{\end{eqnarray}}

 \makeatletter
\def\alt{\mathrel{\mathpalette\gl@align<}}
\def\agt{\mathrel{\mathpalette\gl@align>}}
\def\gl@align#1#2{\lower.6ex\vbox{\baselineskip\z@skip\lineskip\z@
\ialign{$\m@th#1\hfil##\hfil$\crcr#2\crcr\sim\crcr}}} \makeatother

\begin{document}
\begin{flushright}
\end{flushright}
\vspace*{1.0cm}

\begin{center}
\baselineskip 20pt 
{\Large\bf 
Observing Signals of the Bulk Matter RS Model 
\\ through Rare Decays of SUSY Particles
}
\vspace{1cm}

{\large 
Toshifumi Yamada$^{a,b}$
} \vspace{.5cm}

{\baselineskip 20pt \it

$^{a}$ Department of Particles and Nuclear Physics, \\
The Graduate University for Advanced Studies (SOKENDAI), \\

$^{b}$ Institute of Particle and Nuclear Studies, \\ 
High Energy Accelerator Research Organization (KEK),  \\

1-1 Oho, Tsukuba, Ibaraki 305-0801, Japan} 

\vspace{.5cm}

\vspace{1.5cm} {\bf Abstract} \end{center}

The bulk matter Randall-Sundrum (RS) model is a setup
 where Standard Model (SM) matter and gauge fields reside in the bulk of 5D warped spacetime
 while the Higgs field is confined on the IR brane.
The wavefunctions of the 1st and 2nd generation matter particles are localized towards the UV brane
 and those of the 3rd generation towards the IR brane,
 so that the hierarchical structure of the Yukawa couplings arises geometrically
 without hierarchy in fundamental parameters.
This paper discusses observing signals of this model 
 in the case where the Kaluza-Klein scale is far above the collider scale,
 but the model is combined with 5D Minimal SUSY Standard Model (MSSM) 
 and SUSY particles are in the reach of collider experiments.
A general SUSY breaking mass spectrum consistent with 
 the bulk matter RS model is considered:
SUSY breaking sector locates on the IR brane and its effects are mediated to 5D MSSM
 through a hybrid of gravity mediation, gaugino mediation and gauge mediation.
This paper argues that it is possible to
 observe the signals of the bulk matter RS model 
 through rare decays of ``almost SU(2) singlet mass eigenstates"
 that are induced by flavor-violating gravity mediation contributions 
 to matter soft SUSY breaking terms.

\thispagestyle{empty}

\newpage

\addtocounter{page}{-1}
\setcounter{footnote}{0}
\baselineskip 18pt
%

\section{Introduction}

\ \ \ The origin of the fermion mass hierarchy is a long-standing mystery in particle physics.
In the Standard Model (SM), it is explained by the hierarchical Yukawa coupling constants,
 but the hierarchy itself is still introduced by hand.
Many authors have proposed models beyond the SM where the Yukawa coupling hierarchy arises 
 from non-hierarchical couplings of a fundamental theory.
The bulk matter Randall-Sundrum (RS) model \cite{RS, bulk matter} is one of the successful models.
In this model, SM fermions are identified with the zero-modes of 5D Dirac fermions that live in 5D warped spacetime (bulk),
 whereas the Higgs field is confined on 4D spacetime of the infrared (IR) brane.
With non-hierarchical values of 5D Dirac masses, 
 the zero-modes of the 5D fermions can be localized towards either the ultraviolet (UV) brane or the IR brane.
In this way, the geometrical overlap between each zero-mode in the bulk and the Higgs field on the IR brane gains exponential hierarchy,
 which gives rise to the hierarchical structure of the Yukawa coupling constants.
As far as the model gives a natural explanation to the fermion mass hierarchy, 
 neglecting the gauge hierarchy problem,
 it is sufficient that the warp factor be around $\sim m_{e}/m_{t}$, 
 or equivalently the Kaluza-Klein (KK) scale be around $\sim M_{*} \cdot (m_{e}/m_{t})$ 
 ($M_{*}$ indicates the 4D reduced Planck mass) if the 5D Planck scale is the same as that of 4D.
Then the KK modes need not exist at TeV scale.
Futhermore, the physics of flavor has already imposed severe constraints on the mass of KK modes;
 for example, the 1st KK gluon in the bulk matter RS model induces flavor changing neutral current interactions.
 The data on the $K^{0}-\bar{K}^{0}$ mixing require that its mass be larger than $21$ TeV \cite{bound on bulk matter}.

If the KK modes appear only far above the TeV scale, 
 it is impossible to produce them at colliders and confirm the model.
In this paper, I argue that 
 it is possible to observe indirect signatures of the bulk matter RS model at near-future colliders.
I consider the case where the KK scale is at an intermediate scale between Planck and TeV,
 but the 4D effective theory contains ${\cal N}=1$ SUSY which is broken at TeV scale
 and it can be described by Minimal SUSY Standard Model (MSSM).
Hence the SUSY particles are accessible at colliders while the KK modes are beyond their reach.
This is a natural situation because 
 SUSY breaking at TeV scale is necessary to solve the gauge hierarchy problem 
 when the KK scale is at an intermediate scale.
I consider a general SUSY breaking mediation mechanism that is consistent with the bulk matter RS setup,
 in contrast to the paper \cite{gaugino mediation + bulk matter}
 where a simultaneous explanation to the SUSY breaking mediation mechanism and the Yukawa coupling hierarchy
 based on the RS spacetime was pursued.
In the setup of this paper,
 SUSY breaking sector locates on the IR brane and its effects are mediated to MSSM in the bulk
 through contact terms on the IR brane (gravity mediation \cite{gravity mediation}), 
 renormalization group evolutions below the KK scale (gaugino mediation \cite{gaugino mediation})
 and gauge interactions with messenger fields on the IR brane (gauge mediation \cite{gauge mediation}).
(Anomaly mediation contributions \cite{anomaly mediation} are suppressed at least by the warp factor
 compared to gaugino mediation ones and hence are negligible.)

Gravity mediation contributions are the key to observe signals of the bulk matter RS model.
This is because they arise from contact terms on the IR brane in a similar way to the Yukawa coupling constants.
The basic strategy for testing the model is as follows.
Since the zero-modes of matter hypermultiplets (SM fermions are their fermionic components) reside in the bulk, 
 they have contact terms with SUSY breaking sector on the IR brane
 which induce soft SUSY breaking terms through gravity mediation.
The amount of the gravity mediation contribution to each term
 is proportional to the geometrical overlap between the zero-mode fields and the IR brane.
In the bulk matter RS model, the same overlap also gives rise to the hierarchy of the Yukawa coupling constants.
Hence the flavor structure of gravity mediation contributions and the Yukawa coupling hierarchy are related
 and one can predict the former from the latter.
(A similar setup \cite{RS flavorful} was proposed on a different context
 as a 5D realization of ``flavorful supersymmtery" \cite{flavorful}.)
Therefore the model can be tested through 
 the measurement of gravity-mediation-origined soft SUSY breaking terms.
Our task is then to extract gravity mediation contributions from 
 experimentally observable quantities related to soft SUSY breaking terms.
To be a realistic SUSY breaking model, the model must contain dominant flavor-conserving soft SUSY breaking mass terms
 that arise through gaugino mediation and gauge mediation.
On the other hand, gravity mediation intrinsically violates flavor.
Therefore one can indirectly measure the gravity-mediation-origined terms
 through flavor-violating interactions of SUSY matter particles.
One obstacle is that 
 the Yukawa coupling constants themselves induce flavor-violating soft SUSY breaking terms through renormalization group (RG) evolutions,
 as in models with minimal flavor violation.
However, one can distinguish gravity mediation contributions from the RG effects of the Yukawa couplings
 by focusing on SU(2) singlet SUSY matter particles,
 for which flavor violation of gravity mediation contributions 
 can be significantly larger than that of the RG effects.
In this paper, I introduce three promising channels for observing
 signatures of the bulk matter RS model at colliders.
One is that SU(2) singlet smuon mixes with stau through gravity-mediation-origined soft terms
 and one measures the branching ratio of ``almost singlet smuon mass eigenstate"
 decaying into SM tau and another SUSY particle.
Another channel is that SU(2) singlet smuon mixes with selectron
 and one measures the branching ratio of ``almost singlet smuon mass eigenstate"
 decaying into SM electron and another SUSY particle.
The third is that SU(2) singlet scharm mixies with stop
 and one measures the branching ratio of ``almost singlet scharm mass eigenstate"
 decaying into SM top and another SUSY particle.
The bulk matter RS model predicts these branching ratios
 and can be tested through their measurements.
In this setup, one cannot determine the exact values of the contact term couplings of the 5D theory.
I here assume that the contact term couplings for gravity mediation are all $O(1)$,
 and estimate the orders of magnitudes of flavor-violating soft SUSY breaking terms.
Still the bulk matter RS model gives predictions on the magnitudes of the branching ratios of sparticle rare decays.

This paper is organized as follows.
In section 2, I review the bulk matter RS model and combine it with 5D MSSM.
5D disposition of matter wavefunctions are determined so as to reproduce
 the hierarchical structure of the Yukawa coupling constants, Cabbibo-Kobayashi-Maskawa (CKM) matrix and the neutrino mass matrix
 with $O(1)$ couplings of the 5D theory.
In section 3, I derive the SUSY particle mass spectrum with emphasis
 on its flavor-violating sector.
In section 4, I make predictions on the flavor mixings of SUSY particles,
 and compare the bulk matter RS model with other models.
In section 5, I discuss experimental methods to test these predictions.
\\

\section{The Bulk Matter RS Model with 5D MSSM}

\subsection{Setup}

\ \ \ Consider 5D warped spacetime with the metric \cite{RS} : 
\begin{equation}
{\rm d}s^{2} \ = \ e^{-2k \vert y \vert} \eta_{\mu \nu} {\rm d}x^{\mu} {\rm d}x^{\nu} - {\rm d}y^{2} \ , 
\end{equation}
 where $y$ is the 5th dimension compactified on the orbifold $S^{1}/Z_{2} : -\pi R \leq y \leq \pi R $ ,
 and $k$ is the Anti-deSitter curvature that is of the same order as the 5D Planck scale $M_{5}$.
Assuming that the warp factor, $e^{-k R \pi}$, is much smaller than $1$,
 we have the following relation for $k$ and $M_{5}$ : 
\begin{equation}
M_{*}^{2} \ = \ \frac{M_{5}^{3}}{k} ( 1-e^{-2k R \pi} ) \ \simeq \ \frac{M_{5}^{3}}{k} \ ,
\end{equation}
 where $M_{*}$ is the 4D reduced Planck mass,
 which implies $k \sim M_{5} \sim M_{*}$.
The UV brane is put at $y=0$ and the IR brane at $y=\pi R$.
The Planck scale on the UV brane is $M_{5}$,
 while that on the IR brane is $M_{5} e^{-kR\pi}$.
The IR scale, $k e^{-kR\pi} \sim M_{5} e^{-kR\pi}$,
 is assumed to be at an intermediate scale between $M_{*}$ and TeV.
In particular, we assume that it is far above $21$ TeV.
Since the most severe constraint on the IR scale of the bulk matter RS model
 comes from the data on the $K^{0}-\bar{K}^{0}$ mixing, which require it to be larger
 than $21$ TeV \cite{bound on bulk matter},
 my model is free from any constraint on the bulk matter RS model itself.
At the same time, it is hopeless to observe the effects of the Kaluza-Klein excitations
 by near-future experiments.

Consider 5D MSSM \cite{SUSY RS} where the 4D ${\cal N}=1$ Higgs superfields are confined on the IR brane,
 and the 5D ${\cal N}=1$ gauge superfields and matter hypermultiplets live in the bulk.
In the following, we use the 4D superfield formalism extended with the 5th dimension $y$.
We introduce a chiral superfield, $X$, on the IR brane
 whose F-component, $F_{X}$, develops vev to break 4D ${\cal N}=1$ SUSY there.
We consider both cases where there are one to several messenger pairs on the IR brane
 and there is no messenger pair at all.
(It is easy to extend the model to cases where the messengers live in the bulk.)
The gauge symmteries of the messenger pairs are not specified.
SU(2) doublet squark, singlet up-type squark, singlet down-type squark, doublet slepton, singlet charged slepton hypermultiplets
 are denoted by $Q_{i}, U_{i}, D_{i}, L_{i}, E_{i}$, respectively, with $i$ being flavor index.
The up-type Higgs and the down-type Higgs superfields are denoted by $H_{u}, H_{d}$, respectively.

An off-shell 5D ${\cal N}=1$ gauge superfield
 consists of a 5D gauge field $A_{M} \ (M=0,1,2,3,5)$, two 4D Weyl spinors $\lambda_{1}, \lambda_{2}$, a real scalar $\Sigma$,
 a real auxiliary field $D$ and a complex auxiliary field $F$, all of which transform as the adjoint reprsentation of some gauge group.
They combine to form one 4D ${\cal N}=1$ gauge superfield $V$ and one 4D ${\cal N}=1$ chiral superfield $\chi$ that are
\begin{eqnarray*}
V &=& -\theta \sigma^{\mu} \bar{\theta} A_{\mu} - i \bar{\theta}^{2} \theta \lambda_{1} 
+ i \theta^{2} \bar{\theta} \bar{\lambda}_{1} + \frac{1}{2} \bar{\theta}^{2} \theta^{2} D \ ,
\\ \chi &=& \frac{1}{\sqrt{2}} ( \Sigma + i A_{5} ) + \sqrt{2} \theta \lambda_{2} + \theta^{2} F \ .
\end{eqnarray*}
By $Z_{2}$ : $y \rightarrow -y$ symmetry, they transform as 
\begin{eqnarray*}
V \ \rightarrow \ V \ , \ \ \ \ \ \chi \ \rightarrow \ -\chi \ .
\end{eqnarray*}
Assuming the invariance of the theory under the $Z_{2}$ symmtery,
 we obtain the following action for 5D ${\cal N}=1$ gauge superfields:
\begin{eqnarray}
S_{5D \, gauge} &=& \int {\rm d}y \int {\rm d}^{4}x \ e^{-4k \vert y \vert} \ 
\left[ \ \frac{1}{4 (g^{a}_{5})^{2}} \int {\rm d}^{2} \theta e^{k \vert y \vert} \ {\rm tr} \left\{ \ (e^{\frac{3}{2} k \vert y \vert} W^{a \, \alpha}) (e^{\frac{3}{2} k \vert y \vert} W^{a}_{\alpha})  
\ + \ {\rm h.c.} \ \right\} \right. \nonumber
\\ & & \left. \ + \ \frac{1}{(g^{a}_{5})^{2}} \int {\rm d}^{4} \theta e^{2k \vert y \vert} \ 
{\rm tr} \left\{ \ ( \sqrt{2} \partial_{y} + \chi^{a \, \dagger} ) e^{-V}  ( - \sqrt{2} \partial_{y} + \chi^{a} ) e^{V} 
\ - \ (\partial_{y} e^{-V}) (\partial_{y} e^{V}) \right\} \right] \ , \nonumber \\
\end{eqnarray}
 where $a$ labels gauge groups and $W^{a \, \alpha}$ denotes the field strength of $V^{a}$ in 4D flat spacetime.
When the unitary gauge, $A^{a}_{5}=0$, is chosen,
 only $V^{a}$ has a massless mode in 4D picture.
This mode has no dependence on $y$ and will be written as $V_{0}(x,\theta,\bar{\theta})$.

A 5D ${\cal N}=1$ hypermultiplet is expressed in terms of two 4D ${\cal N}=1$ chiral superfields $\Phi, \Phi^{c}$
 that are in conjugate representaions of some gauge group.
We assume that the former is $Z_{2}$-even and the latter $Z_{2}$-odd.
Taking the basis of diagonal bulk masses, we have the following action for 5D ${\cal N}=1$ hypermultiplets:
\begin{eqnarray}
S_{5D \, chiral} &=& \int {\rm d}y \int {\rm d}^{4}x e^{-4k \vert y \vert} \ \left[ \ \int {\rm d}^{4} \theta e^{2k \vert y \vert} \ \right.
 ( \Phi_{i}^{\dagger} e^{-V} \Phi_{i} \ + \ \Phi_{i}^{c} e^{V} \Phi_{i}^{c \, \dagger} ) \nonumber
\\ & & \left. \ + \ \int {\rm d}^{2} \theta e^{k \vert y \vert} \ \Phi_{i}^{c} \{ \partial_{y} - \chi/\sqrt{2} - (3/2-c_{i}) k \} \Phi_{i} \ + \ {\rm h.c.} \ \right] \ ,
\end{eqnarray}
 where $i$ is a flavor index and $c_{i}$ denotes the 5D bulk mass in unit of AdS curvature $k$.
Only $\Phi_{i}$ has a massless mode in 4D picture, which will be written as \ $\phi_{i}(x, \theta) e^{(3/2-c_{i}) k \vert y \vert}$.

We write down the 4D effective action for the fields in the bulk in terms of the massless modes:
\begin{eqnarray}
S_{4D \, eff.} &=& \int {\rm d}^{4}x \ \left[ \ \frac{2 \pi R}{4 g_{5}^{a \, 2}} \int {\rm d}^{2}\theta \ W^{a \alpha} W^{a}_{\alpha} \ + \ {\rm h.c.} \right. \nonumber 
\\ &+& \left. \int {\rm d}^{4}\theta \ 2 \frac{ e^{(1-2c_{i})kR\pi}-1 }{(1-2c_{i})k} \ \phi_{i}^{\dagger} \ e^{-V} \phi_{i} \ \right] \ , \nonumber \\ 
\end{eqnarray}
 where the dimensionful 5D gauge coupling, $g_{5}^{a}$, is connected to 4D gauge coupling $g_{4}^{a}$ by the relation:
 $g_{5}^{a} = \sqrt{2\pi R} g_{4}^{a}$.
$\phi_{i}$ represents the zero-mode of each of $Q_{i}, U_{i}, D_{i}, L_{i}, E_{i}$.

We also introduce an IR-brane-localized action.
Below are the parts of the action relevant to the topic of this paper.

MSSM term:
\begin{eqnarray}
S_{IR} &\supset& \int {\rm d}^{4}x \ \left[ \int {\rm d}^{4}\theta \ e^{-2kR\pi} \ \left\{ \ H_{u}^{\dagger} e^{-V} H_{u} \ + \ H_{d}^{\dagger} e^{-V} H_{d} \ \right\} \right. \nonumber
\\ &+& \int {\rm d}^{2}\theta \ e^{-3kR\pi} \ \left\{ \ e^{(3-c_{i}-c_{j})kR\pi} \ \frac{(y_{u})_{ij}}{M_{5}} H_{u} U_{i} Q_{j} 
\ + \ e^{(3-c_{k}-c_{l})kR\pi} \ \frac{(y_{d})_{kl}}{M_{5}} H_{d} D_{k} Q_{l} \ \right\} \ + \ {\rm h.c.} \nonumber
\\ &+& \left. \int {\rm d}^{2}\theta \ e^{-3kR\pi} \ e^{(3-c_{m}-c_{n})kR\pi} \ \frac{(y_{e})_{mn}}{M_{5}} H_{d} E_{m} L_{n} \ + \ {\rm h.c.} \ \right] \ .
\end{eqnarray}

Gaugino mass term:
\begin{eqnarray}
S_{IR} &\supset& \int {\rm d}^{4}x \ \left[ \int {\rm d}^{2}\theta \ d_{a} \frac{X}{M_{5}} W^{a \, \alpha} W^{a}_{\alpha} \ + \ {\rm h.c.} \right] \ .
\end{eqnarray}

Matter soft SUSY breaking mass term:
\begin{eqnarray}
S_{IR} &\supset& \int {\rm d}^{4}x \ \left[ \int {\rm d}^{4}\theta \ e^{-2kR\pi} \ e^{(3-c_{i}-c_{j})kR\pi} \ 
\left\{ \ d_{Q1 \, ij} \frac{X+X^{\dagger}}{M^{2}_{5}} \ Q_{i}^{\dagger} Q_{j} \ + \ d_{Q2 \, ij} \frac{X^{\dagger} X}{M^{3}_{5}} \ Q_{i}^{\dagger} Q_{j} \ \right\} \right] \nonumber
\\ & + & ( \ Q \ \rightarrow \ U, \ D, \ L, \ E \ ) \ .
\end{eqnarray}

A-term-generating term:
\begin{eqnarray}
S_{IR} &\supset& \int {\rm d}^{4}x \ \left[ \int {\rm d}^{2}\theta \ e^{-3kR\pi} \ \left\{ \ e^{(3-c_{i}-c_{j})kR\pi} \ \frac{(a_{u})_{ij}}{M_{5}^{2}} X H_{u} U_{i} Q_{j} 
\ + \ e^{(3-c_{k}-c_{l})kR\pi} \ \frac{(a_{d})_{kl}}{M_{5}^{2}} X H_{d} D_{k} Q_{l} \ \right. \right. \nonumber
\\ & + & \left. \left. e^{(3-c_{m}-c_{n})kR\pi} \ \frac{(a_{e})_{mn}}{M_{5}^{2}} X H_{d} E_{m} L_{n} \ \right\} \ + \ {\rm h.c.} \ \right] \ .
\end{eqnarray}

Messenger term:
\begin{eqnarray}
S_{IR} &\supset&  \sum_{I} \ \int {\rm d}^{4}x \ \left[ \int {\rm d}^{4}\theta \ e^{-2kR\pi} \
\{ \ \Xi_{I}^{\dagger} e^{-V} \Xi_{I} + \bar{\Xi}_{I}^{\dagger} e^{V} \bar{\Xi}_{I} \ \} \right. \nonumber
\\ & + & \left. \int {\rm d}^{2}\theta \ e^{-3kR\pi} \ \{ \ M_{mess \, I} \ \Xi_{I} \bar{\Xi}_{I} \ + \ 
\lambda_{mess \, I} \ X \ \Xi_{I} \bar{\Xi}_{I} \ \} \ + \ {\rm h.c.} \ \right] \ ,
\end{eqnarray}
 where $M_{mess \, I}$ indicates the SUSY conserving mass for the messenger pair $\Xi_{I}, \bar{\Xi}_{I}$.
Note that we did not necessarily assume the existence of messengers.
In that case, only gaugino mediation gives rise to flavor-conserving soft masses, 
 as is realized in the model \cite{gaugino mediation + bulk matter}.

In addition, the terms for the Higgs superfields exist on the IR brane.
We simply assume that $\mu$-term and $B\mu$-term are somehow derived at TeV scale.

We normalize $X, \ H_{u}, \ H_{d}, \ Q_{i}, \ U_{i}, \ D_{i}, \ L_{i}, \ E_{i}, \ \Xi_{I}, \ \bar{\Xi}_{I}$ 
 to make their kinetic terms in the 4D effective theory canonical. 
This is done by the following rescaling:
\begin{eqnarray}
X \ \rightarrow \ \tilde{X}=e^{-kR\pi}X,
\ \ H_{u} \ \rightarrow \ \tilde{H}_{u}=e^{-kR\pi} H_{u}, 
\ \ H_{d} \ \rightarrow \ \tilde{H}_{d}=e^{-kR\pi} H_{d}, \nonumber \\
\phi_{i} \ \rightarrow \ \tilde{\phi}_{i}=\sqrt{ 2 \frac{e^{(1-2c_{i})kR\pi}-1}{(1-2c_{i})k} } \ \phi_{i}, \nonumber \\
\Xi_{I} \ \rightarrow \ \tilde{\Xi}_{I}=e^{-kR\pi} \Xi_{I},
\ \ \bar{\Xi}_{I} \ \rightarrow \ \tilde{\bar{\Xi}}_{I}=e^{-kR\pi} \bar{\Xi}_{I}.
\end{eqnarray}

Then the MSSM term, the gaugino mass term, the matter soft SUSY breaking mass term,
 the A-term-generating term and the messenger term become as follows:
\begin{eqnarray}
S_{IR} &\supset& \int {\rm d}^{4}x \ \left[ \ \int {\rm d}^{4}\theta \ \left\{ \ \tilde{H}_{u}^{\dagger} e^{-V} \tilde{H}_{u} 
\ + \ \tilde{H}_{d}^{\dagger} e^{-V} \tilde{H}_{d} \ \right\} \right. \nonumber
\\ &+& \int {\rm d}^{2}\theta \ \left\{ \ \sqrt{ \frac{1-2c_{i}}{2\{1-e^{-(1-2c_{i})kR\pi}\}} } \sqrt{ \frac{1-2c_{j}}{2\{1-e^{-(1-2c_{j})kR\pi}\}} } 
\frac{k}{M_{5}} (y_{u})_{ij} \ \tilde{H}_{u} \tilde{U}_{i} \tilde{Q}_{j} \right. \nonumber
\\ &+& \sqrt{ \frac{1-2c_{k}}{2\{1-e^{-(1-2c_{k})kR\pi}\}} } \sqrt{ \frac{1-2c_{l}}{2\{1-e^{-(1-2c_{l})kR\pi}\}} } 
\frac{k}{M_{5}} (y_{d})_{kl} \ \tilde{H}_{d} \tilde{D}_{k} \tilde{Q}_{l} \nonumber
\\ &+& \left. \left. \sqrt{ \frac{1-2c_{m}}{2\{1-e^{-(1-2c_{m})kR\pi}\}} } \sqrt{ \frac{1-2c_{n}}{2\{1-e^{-(1-2c_{n})kR\pi}\}} }
\frac{k}{M_{5}} (y_{e})_{mn} \ \tilde{H}_{d} \tilde{E}_{m} \tilde{L}_{n} \ \right\} \ + \ {\rm h.c.} \right] .
\end{eqnarray}

The gaugino mass term will be
\begin{eqnarray}
S_{IR} &\supset& \int {\rm d}^{4}x \ \left[ \int {\rm d}^{2}\theta \ d_{a} \frac{\tilde{X}}{M_{5}e^{-kR\pi}} W^{a \, \alpha} W^{a}_{\alpha} \ + \ {\rm h.c.} \right] \ .
\end{eqnarray}

The matter soft mass term will be
\begin{eqnarray}
S_{IR} &\supset& \int {\rm d}^{4}x \ \left[ \int {\rm d}^{4}\theta \ \sqrt{ \frac{1-2c_{i}}{2\{1-e^{-(1-2c_{i})kR\pi}\}} } \sqrt{ \frac{1-2c_{j}}{2\{1-e^{-(1-2c_{j})kR\pi}\}} } \ \frac{k}{M_{5}} \ \times \right. \nonumber
\\
& & \ \ \left. \left\{ \ d_{Q1 \, ij} \frac{\tilde{X}+\tilde{X}^{\dagger}}{M_{5}e^{-kR\pi}} \ \tilde{Q}_{i}^{\dagger} \tilde{Q}_{j} \ + \ 
d_{Q2 \, ij} \frac{\tilde{X}^{\dagger} \tilde{X}}{M^{2}_{5} e^{-2kR\pi}} \ \tilde{Q}_{i}^{\dagger} \tilde{Q}_{j} \  \right\} \right] \nonumber
\\ &+& ( \ \tilde{Q} \ \rightarrow \ \tilde{U}, \ \tilde{D}, \ \tilde{L}, \ \tilde{E} \ ) \ .
\end{eqnarray}

The A-term-generating term will be
\begin{eqnarray}
S_{IR} &\supset& \int {\rm d}^{4}x \ \left[ \int {\rm d}^{2}\theta \ \left\{ \ \sqrt{ \frac{1-2c_{i}}{2\{1-e^{-(1-2c_{i})kR\pi}\}} } \sqrt{ \frac{1-2c_{j}}{2\{1-e^{-(1-2c_{j})kR\pi}\}} } 
\frac{k}{M_{5}} \frac{(a_{u})_{ij}}{M_{5} e^{-kR\pi}} \ \tilde{X} \tilde{H}_{u} \tilde{U}_{i} \tilde{Q}_{j} \right. \right. \nonumber
\\ &+& \sqrt{ \frac{1-2c_{k}}{2\{1-e^{-(1-2c_{k})kR\pi}\}} } \sqrt{ \frac{1-2c_{l}}{2\{1-e^{-(1-2c_{l})kR\pi}\}} } 
\frac{k}{M_{5}} \frac{(a_{d})_{kl}}{M_{5} e^{-kR\pi}} \ \tilde{X} \tilde{H}_{d} \tilde{D}_{k} \tilde{Q}_{l} \nonumber
\\ &+& \left. \left. \sqrt{ \frac{1-2c_{m}}{2\{1-e^{-(1-2c_{m})kR\pi}\}} } \sqrt{ \frac{1-2c_{n}}{2\{1-e^{-(1-2c_{n})kR\pi}\}} }
\frac{k}{M_{5}} \frac{(a_{e})_{mn}}{M_{5} e^{-kR\pi}} \ \tilde{X} \tilde{H}_{d} \tilde{E}_{m} \tilde{L}_{n} \ \right\} \ + \ {\rm h.c.} \ \right] \ . \nonumber \\
\end{eqnarray}

The messenger term will be
\begin{eqnarray}
S_{IR} &\supset&  \sum_{I} \ \int {\rm d}^{4}x \ \left[ \int {\rm d}^{4}\theta \
\{ \ \tilde{\Xi}_{I}^{\dagger} e^{-V} \tilde{\Xi}_{I} + \tilde{\bar{\Xi}}_{I}^{\dagger} e^{V} \tilde{\bar{\Xi}}_{I} \ \} \right. \nonumber
\\ & + & \left. \int {\rm d}^{2}\theta \ \{ \ M_{mess \, I}e^{-kR\pi} \ \tilde{\Xi}_{I} \tilde{\bar{\Xi}}_{I} \ + \ 
\lambda_{mess \, I} \tilde{X} \ \tilde{\Xi}_{I} \tilde{\bar{\Xi}}_{I} \ \} \ + \ {\rm h.c.} \ \right] \ .
\end{eqnarray}

We introduce light neutrino masses by writing an IR-scale-suppressed higher dimensional operators 
 or by adopting the seesaw mechanism \cite{seesaw}.
In either case, we have the following term for light neutrino masses:
\begin{eqnarray}
S_{IR} &\supset& \int {\rm d}^{4}x \int {\rm d}^{2}\theta \ e^{-3kR\pi} \ e^{(3-c_{p}-c_{q})kR\pi} \ 
(Y_{\nu})_{pq} \frac{L_{p} H_{u} L_{q} H_{u}}{M_{seesaw}} \ + \ {\rm h.c.} \ \nonumber
\\ &=& \int {\rm d}^{4}x \int {\rm d}^{2}\theta \ \sqrt{ \frac{1-2c_{p}}{2\{1-e^{-(1-2c_{p})kR\pi}\}} } \sqrt{ \frac{1-2c_{q}}{2\{1-e^{-(1-2c_{q})kR\pi}\}} } \ 
(Y_{\nu})_{pq} \ \frac{\tilde{L}_{p} \tilde{H}_{u} \tilde{L}_{q} \tilde{H}_{u}}{M_{seesaw} e^{-kR\pi}} \ + \ {\rm h.c.} \ , \nonumber \\
\end{eqnarray}
 where $M_{seesaw}$ indicates the mass scale relevant to the light neutrino mass.
To maintain the generality of the model, we hereafter consider cases with SU(2) singlet neutrinos.
Their Majorana masses in the \textit{4D effective} theory are assumed to be around 
 a common scale, denoted by $M_{Maj}$,
 which is lower than $M_{mess \, I}e^{-kR\pi}$ or $M_{5}e^{-kR\pi}$.
The results of this paper can be extended to cases without singlet neutrinos
 by dropping terms containing $M_{Maj}$.
\\

Now the MSSM Yukawa coupling constants are expressed as :
\begin{eqnarray}
(Y_{u})_{ij} &=& \sqrt{ \frac{1-2c_{i}}{2\{1-e^{-(1-2c_{i})kR\pi}\}} } \sqrt{ \frac{1-2c_{j}}{2\{1-e^{-(1-2c_{j})kR\pi}\}} }
\frac{k}{M_{5}} (y_{u})_{ij} \ , \nonumber
\\ 
(Y_{d})_{kl} &=& \sqrt{ \frac{1-2c_{k}}{2\{1-e^{-(1-2c_{k})kR\pi}\}} } \sqrt{ \frac{1-2c_{l}}{2\{1-e^{-(1-2c_{l})kR\pi}\}} }
\frac{k}{M_{5}} (y_{d})_{kl} \ , \nonumber
\\
(Y_{e})_{mn} &=& \sqrt{ \frac{1-2c_{m}}{2\{1-e^{-(1-2c_{m})kR\pi}\}} } \sqrt{ \frac{1-2c_{n}}{2\{1-e^{-(1-2c_{n})kR\pi}\}} }
\frac{k}{M_{5}} (y_{e})_{mn} \ ,
\end{eqnarray}
 and the neutrino mass matrix $m_{\nu}$ is given by :
\begin{eqnarray}
(m_{\nu})_{pq} &=& \sqrt{ \frac{1-2c_{p}}{2\{1-e^{-(1-2c_{p})kR\pi}\}} } \sqrt{ \frac{1-2c_{q}}{2\{1-e^{-(1-2c_{q})kR\pi}\}} }
\ (Y_{\nu})_{pq} \ \frac{v_{u}^{2}}{M_{seesaw}e^{-kR\pi}} \ .
\end{eqnarray}
The geometrical factor $\sqrt{ (1-2c) \ / \ (2\{1-e^{-(1-2c)kR\pi}\}) }$ 
 generates hierarchical couplings without fundamental hierarchy \cite{bulk matter};
 for $c < 1/2$, it is approximated by $\sqrt{ 1/2 - c }$ and is $O(1)$, 
 whereas for $c > 1/2$, it is approximated by $\sqrt{ c - 1/2 } \ e^{-(c-1/2)kR\pi}$ and is exponentially suppressed.
Note that this factor cannot be larger than $O(1)$.
We assume that the components of 5D coupling matrices $y_{u}, \ y_{d}, \ y_{e}$ are all $O(1)$
 and that the hierarchy of the Yukawa coupling constants stems solely from the following terms:
\begin{eqnarray*}
\sqrt{ \frac{1-2c_{i}}{2\{1-e^{-(1-2c_{i})kR\pi}\}} } \sqrt{ \frac{1-2c_{j}}{2\{1-e^{-(1-2c_{j})kR\pi}\}} } \ .
\end{eqnarray*}
This is how the bulk matter RS model explains the Yukawa coupling hierarchy.

We further assume that the components of $Y_{\nu}$ are $O(1)$.
Note that $Y_{\nu}$ arises by integrating out singlet Majorana neutrinos.
If the components of 5D neutrino Dirac coupling are $O(1)$,
 it is possible to take the value of $M_{seesaw}$
 such that $(Y_{\nu})_{ij} \sim O(1)$ holds,
 regardless of the 5D disposition of singlet neutrino fields and the flavor structure of the Majorana mass term.
Hence this is a natural assumption in the bulk matter RS model, in which all 5D couplings are considered $O(1)$.
With this assumption, the hierarchy of the light neutrino mass matrix (19) arises only from
 the geometrical factors of SU(2) doublet lepton fields.
\\

We hereafter use the following notations:
\begin{equation}
\alpha_{i} \ \equiv \ \sqrt{ \frac{1-2c_{q \, i}}{2\{1-e^{-(1-2c_{q \, i})kR\pi}\}} } \ \ \ \ \ {\rm with} \ \ i=1,2,3
\end{equation}
 for SU(2) doublet quark superfields with flavor index $i$,
 and $\beta_{i}, \ \gamma_{i}, \ \delta_{i}, \ \epsilon_{i}$
 for SU(2) singlet up-type quark, singlet down-type quark, doublet lepton and singlet charged lepton, respectively.
Then the hierarchical structures of the up-type quark Yukawa matrix $Y_{u}$, the down-type Yukawa matrix $Y_{d}$ and the charged lepton Yukawa matrix $Y_{e}$ 
 (in the basis of diagonal 5D bulk masses) are expressed as:
\begin{equation}
(Y_{u})_{ij} \ \sim \ \beta_{i}\alpha_{j} \ , \ \ \ (Y_{d})_{ij} \ \sim \ \gamma_{i}\alpha_{j} \ , \ \ \ (Y_{e})_{ij} \ \sim \ \epsilon_{i}\delta_{j} \ ,
\end{equation}
 and that of the neutrino mass matrix $m_{\nu}$ is expressed as:
\begin{equation}
(m_{\nu})_{ij} \ \sim \ \delta_{i} \delta_{j} \ \frac{v_{u}^{2}}{M_{seesaw}e^{-kR\pi}} \ .
\end{equation}
\\

\subsection{Determination of the Geometrical Factors}

\ \ \ The magnitudes of the geometrical factors, $\alpha_{i}, \beta_{i}, \gamma_{i}, \delta_{i}, \epsilon_{i}$,
 can be almost determined by the data on SM fermion masses, CKM matrix and neutrino oscillations.
The sole exception is the absolute scale of $\delta_{i}$s, 
 of which we only know the relative scales between different flavors.
In this subsection, we will estimate these factors.
The values that correspond to the model must be given at the KK scale, $M_{5}e^{-kR\pi}$, where the 5D theory is connected to the 4D effective theory.
However, as is seen from \cite{scale dependence of}, RG evolutions
 change the Yukawa coupling constants by at most $2$
 and the CKM matrix components by at most $1.2$ through evolving from $\sim 10^{15}$ GeV to the electroweak scale.
Also the neutrino mass matrix is affected only by $O(1)$ through RG evolutions \cite{RGE for neutrinos}.
Therefore we may estimate the magnitudes of $\alpha_{i}, \beta_{i}, \gamma_{i}, \delta_{i}, \epsilon_{i}$ 
 from the data at low energies.

We first derive the model's predictions 
 on the eigenvalues of the Yukawa coupling matrices and the components of CKM matrix.
Let us diagonalize the Yukawa matrices:
\begin{eqnarray*}
V_{u} Y_{u} U_{u}^{\dagger} &=& {\rm diag} \ ,
\\ V_{d} Y_{d} U_{d}^{\dagger} &=& {\rm diag} \ ,
\\ V_{e} Y_{e} U_{e}^{\dagger} &=& {\rm diag} \ .
\end{eqnarray*}
For successful diagonalization of the hierarchical Yukawa matrices, 
 the unitary matrices, \\ $U_{u}, \ U_{d}, \ V_{u}, \ V_{d}, \ U_{e}, \ V_{e}, $ need to have the following structure: 
\begin{eqnarray}
U_{u} \ \sim \ U_{d} \ \sim \ \left(
\begin{array}{ccc}
1 & 0 & 0 \\
\alpha_{1}/\alpha_{2} & 1 & 0 \\
\alpha_{1}/\alpha_{3} & \alpha_{2}/\alpha_{3} & 1
\end{array}
\right) \ , \ \ \ 
V_{u} \ \sim \ ({\rm \alpha \rightarrow \beta}) \ , \ \ \ V_{d} \ \sim \ ({\rm \alpha \rightarrow \gamma}) \ , \nonumber
\\ U_{e} \ \sim \ ({\rm \alpha \rightarrow \delta}) \ , \ \ \ V_{e} \ \sim \ ({\rm \alpha \rightarrow \epsilon})
\end{eqnarray}
 which leads to
\begin{eqnarray}
V_{u} Y_{u} U_{u}^{\dagger} &\sim& {\rm diag} \ ( \ \beta_{1}\alpha_{1}, \ \beta_{2}\alpha_{2}, \ \beta_{3}\alpha_{3} \ ) \ , \nonumber
\\ V_{d} Y_{d} U_{d}^{\dagger} &\sim& {\rm diag} \ ( \ \gamma_{1}\alpha_{1}, \ \gamma_{2}\alpha_{2}, \ \gamma_{3}\alpha_{3} \ ) \ , \nonumber
\\ V_{e} Y_{e} U_{e}^{\dagger} &\sim& {\rm diag} \ ( \ \epsilon_{1}\delta_{1}, \ \epsilon_{2}\delta_{2}, \ \epsilon_{3}\delta_{3} \ ) \ .
\end{eqnarray}
The hierarchical structure of CKM matrix $U_{CKM}$ is given by 
\begin{eqnarray}
U_{CKM} &=& U_{u} U_{d}^{\dagger} \ \sim \ \left(
\begin{array}{ccc}
1 & \alpha_{1}/\alpha_{2} & \alpha_{1}/\alpha_{3} \\
\alpha_{1}/\alpha_{2} & 1 & \alpha_{2}/\alpha_{3} \\
\alpha_{1}/\alpha_{3} & \alpha_{2}/\alpha_{3} & 1
\end{array}
\right) \ .
\end{eqnarray}

We next list the experimental data on CKM matrix and the neutrino mass matrix.
The absolute values of the CKM matrix components, $\vert U_{CKM} \vert$, at the electroweak scale has been measured to be \cite{pdg}
\begin{eqnarray*}
& & \vert U_{CKM} [ M_{W} ] \vert \\ &=& \left(
\begin{array}{ccc}
0.97419 \pm 0.00022 & 0.2257 \pm 0.0010 & 0.00359 \pm 0.00016 \\
0.2256 \pm 0.0010 & 0.97334 \pm 0.00023 & 0.0415 + 0.0010 -0.0011 \\
0.00874 + 0.00026 - 0.00037 & 0.0407 \pm 0.0010 & 0.999133 + 0.000044 - 0.000043
\end{array}
\right) \ .
\end{eqnarray*}
We approximate this matrix by the following formula:
\begin{equation}
\vert U_{CKM} \vert \ \simeq \ \left(
\begin{array}{ccc}
1 & \lambda & \lambda^{3} \\
\lambda & 1 & \lambda^{2} \\
\lambda^{3} & \lambda^{2} & 1
\end{array}
\right) \ \ {\rm with} \ \ \lambda = 0.22 \ .
\end{equation}
To discuss the neutrino mass matrix, we adopt the tri-bi-maximal mixing matrix:
\begin{eqnarray*}
U_{MNS} &=& \left(
\begin{array}{ccc}
\sqrt{\frac{2}{3}} & \sqrt{\frac{1}{3}} & 0 \\
\sqrt{\frac{1}{6}} & \sqrt{\frac{1}{3}} & \sqrt{\frac{1}{2}} \\
\sqrt{\frac{1}{6}} & \sqrt{\frac{1}{3}} & \sqrt{\frac{1}{2}}
\end{array}
\right)
\end{eqnarray*}
 and the following data on neutrino mass squared differences \cite{pdg}:
\begin{eqnarray*}
\Delta m_{21}^{2} &=& 7.59 \pm 0.20 \times 10^{-5} \ {\rm eV}^{2}, \ \ \ \ \ 
\vert \Delta m_{32}^{2} \vert \ = \  2.43 \pm 0.13 \times 10^{-3} \ {\rm eV}^{2}.
\end{eqnarray*}
We assume that the mass of the lightest neutrino is negligible.
Then the neutrino mass matrix, $ U_{MNS} \ {\rm diag} \ ( \ m_{\nu 1}, \ m_{\nu 2}, \ m_{\nu 3} \ ) \ U_{MNS}^{\dagger} $ ,
 is given by
\begin{eqnarray}
U_{MNS} \ {\rm diag} \ ( \ m_{\nu 1}, \ m_{\nu 2}, \ m_{\nu 3} \ ) \ U_{MNS}^{\dagger} &=& \left(
\begin{array}{ccc}
0.29 & 0.29 & 0.29 \\
0.29 & 2.8 & -2.2 \\
0.29 & -2.2 & 2.8
\end{array}
\right) \times 10^{-11} \ {\rm GeV} \nonumber \\ 
{\rm for \ normal \ hierarchy} \ , \nonumber \\
\\
U_{MNS} \ {\rm diag} \ ( \ m_{\nu 1}, \ m_{\nu 2}, \ m_{\nu 3} \ ) \ U_{MNS}^{\dagger} &=& \left(
\begin{array}{ccc}
4.9 & 0.026 & 0.026 \\
0.026 & 2.5 & 2.5 \\
0.026 & 2.5 & 2.5
\end{array}
\right) \times 10^{-11} \ {\rm GeV} \nonumber \\
{\rm for \ inverted \ hierarchy} \ . \nonumber \\
\end{eqnarray}

We now compare the predictions of the model with the data
 and estimate the magnitudes of $\alpha_{i}, \ \beta_{i}, \ \gamma_{i}, \ \delta_{i}, \ \epsilon_{i}$ .
For Yukawa eigenvalues, we simply have 
\begin{eqnarray}
\beta_{1}\alpha_{1} &\sim& m_{u}/v \sin \beta \ , \ \ \ \beta_{2}\alpha_{2} \ \sim \ m_{c}/v \sin \beta \ , \ \ \ \beta_{3}\alpha_{3} \ \sim \ m_{t}/v \sin \beta \ ,
\\ \gamma_{1}\alpha_{1} &\sim& m_{d}/v \cos \beta \ , \ \ \ \gamma_{2}\alpha_{2} \ \sim \ m_{s}/v \cos \beta \ , \ \ \ \gamma_{3}\alpha_{3} \ \sim \ m_{b}/v \cos \beta \ ,
\\ \epsilon_{1}\delta_{1} &\sim& m_{e}/v \cos \beta \ , \ \ \ \epsilon_{2}\delta_{2} \ \sim \ m_{\mu}/v \cos \beta \ , \ \ \ \epsilon_{3}\delta_{3} \ \sim \ m_{\tau}/v \cos \beta \ ,
\end{eqnarray}
 where the mass values can be approximated by their pole values.
Since the top Yukawa coupling is $\sim 1$, we have $\alpha_{3}\beta_{3} \sim 1$,
 which leads to 
\begin{equation}
\alpha_{3} \ \sim \ 1 \ , \ \ \beta_{3} \ \sim \ 1 \ .
\end{equation}
Comparing (25) with (26), we find that putting
\begin{equation}
\alpha_{1} \ \sim \ \lambda^{3} \ , \ \ \alpha_{2} \ \sim \ \lambda^{2}
\end{equation}
 works.
We then have 
\begin{eqnarray}
\beta_{1} & \sim & \lambda^{-3} \ m_{u}/v \sin \beta \ , \ \ \beta_{2} \ \sim \ \lambda^{-2} \ m_{c}/v \sin \beta\ ,
\\ \gamma_{1} & \sim & \lambda^{-3} \ m_{d}/v \cos \beta \ , \ \ \gamma_{2} \ \sim \ \lambda^{-2} \ m_{s}/v \cos \beta \ ,
\ \ \gamma_{3} \ \sim \ m_{b}/v \cos \beta \ .
\end{eqnarray}
Next compare the matrix (22) with the neutrino mass matrix.
For normal hierarchy case, it is possible to reproduce the hierachical structure of the neutrino mass matrix by assuming
 the relation:
\begin{eqnarray}
3 \delta_{1} &\sim& \delta_{2} \ \sim \ \delta_{3} \ ,
\end{eqnarray}
 and the ratio up to $3$ among the components of 5D coupling $Y_{\nu}$.
In contrast, for inverted hierarchy case, 
 $\sim 200$ ratio is required among the 5D coupling components no matter how we choose $\delta_{i}$s,
 which makes it difficult to naturally explain the hierarchy of the neutrino mass matrix.
The situation gets worse if we consider non-negiligible mass of the lightest neutrino.
In conclusion, the bulk matter RS model favors the normal hierarchy of neutrino masses
 and implies the relation (34) for $\delta_{i}$s.
We estimate $\epsilon_{i}$s assuming the relation (36); 
 we obtain
\begin{equation}
\epsilon_{1} \ \sim \ 3 \ \delta_{3}^{-1} \ m_{e}/v \cos \beta \ , \ \ \epsilon_{2} \ \sim \ \delta_{3}^{-1} \ m_{\mu}/v \cos \beta \ , \ \ 
\epsilon_{3} \ \sim \ \delta_{3}^{-1} \ m_{\tau}/v \cos \beta \ .
\end{equation}
The magnitude of $\delta_{3}$ is a free parameter because we do not specify the scale of $M_{seesaw}$.
\\

\section{Flavor-Violating Soft SUSY Breaking Terms}

\ \ \ In this model, flavor-conserving soft SUSY breaking terms arise 
 from RG contributions of gaugino masses below the KK scale (gaugino mediation)
 and gauge interactions with messenger superfields (gauge mediation).
On the other hand, flavor-violating terms 
 arise from contact interactions with SUSY breaking sector on the IR brane (gravity mediation)
 and RG contributions of the Yukawa couplings.
Of particular importance are the gravity mediation contributions, 
 which have a flavor structure unique to the bulk matter RS model.
In this section, we separately estimate the gravity mediation contributions
 and the Yukawa coupling contributions to flavor-violaing soft SUSY breaking terms.

Remember that there are two scales of soft SUSY breaking terms,
 namely gravity mediation scale and gauge mediation scale.
Assuming that the messenger masses and couplings are around the same orders,
 we define the following two mass parameters:
\begin{eqnarray}
M_{grav} &\equiv& \frac{ \vert <F_{\tilde{X}}> \vert }{ M_{5}e^{-kR\pi} } \ , \\
M_{gauge} &\equiv& \frac{1}{16\pi^{2}} \frac{ \lambda_{mess} \ \vert <F_{\tilde{X}}> \vert }{ M_{mess}e^{-kR\pi} } \ ,
\end{eqnarray}
 where $M_{mess}$ represents the typical scale of the SUSY conserving messenger masses $M_{mess \, I}$,
 and $\lambda_{mess}$ the typical scale of the messenger couplings to SUSY breaking sector $\lambda_{mess \, I}$.
Note that Yukawa RG contributions to flavor-violating terms depend on both $M_{grav}$ and $M_{gauge}$,
 whereas gravity mediation contributions do only on $M_{grav}$.
\\

\subsection{Flavor-Violating Gravity Mediation Contributions}

\ \ \ Let us estimate the magnitudes of gravity mediation contributions in the bulk matter RS model.

From (14), we obtain the following formulae for gravity-mediation-origined soft SUSY breaking matter mass terms
 at the scale $M_{5}e^{-kR\pi}$:

For SU(2) doublet squarks, we have
\begin{eqnarray}
(m_{Q}^{2})_{ij} &=& (-d_{Q2ij}+d_{Q1ij}^{2}) \ \frac{k}{M_{5}} \ \alpha_{i} \alpha_{j} \ M_{grav}^{2} \ .
\end{eqnarray}
By substituting $(U, \beta), \ (D, \gamma), \ (L, \delta), \ (E, \epsilon)$
 into $(Q, \ \alpha)$ in the above formula,
 we obtain similar expressions for SU(2) singlet up-type squarks, down-type squarks, 
 SU(2) doublet sleptons and singlet charged sleptons.

Assuming that the 5D couplings $d_{*2ij}$, $d_{*1ij}$ are $O(1)$,
 we obtain the following estimates on the magnitudes at the scale $M_{5}e^{-kR\pi}$:
\begin{eqnarray}
(m_{Q}^{2})_{ij} &\sim& \alpha_{i} \alpha_{j} \ M_{grav}^{2} \ .
\end{eqnarray}
 and similar formulae with $(U, \beta), \ (D, \gamma), \ (L, \delta), \ (E, \epsilon)$
 replacing $(Q, \ \alpha)$ in the above formula.

Next we estimate the magnitudes of the A-terms that are induced by gravity mediation.
The terms (15) directly contribute to the A-terms.
Furthermore, since the Higgs superfields can couple to SUSY breaking sector in the following way:
\begin{eqnarray*}
\int {\rm d}^{4}\theta \ \left[ \ 
d_{uA} \frac{\tilde{X}}{M_{5}e^{-kR\pi}} \tilde{H}_{u}^{\dagger} \tilde{H}_{u} 
\ + \ d_{dA} \frac{\tilde{X}}{M_{5}e^{-kR\pi}} \tilde{H}_{d}^{\dagger} \tilde{H}_{d}
\ + \ {\rm h.c.} \ \right] \ ,
\end{eqnarray*}
 the A-terms also arise from the Higgs F-terms via (12).
Hence gravity-mediation-origined A-terms at the scale $M_{5}e^{-kR\pi}$ are given by
\begin{eqnarray}
A_{uij} &=& -d_{uA} \ (y_{u})_{ij} \ \beta_{i} \alpha_{j} \ \frac{k}{M_{5}} \ M_{grav}
\ + \ (a_{u})_{ij} \ \beta_{i} \alpha_{j} \ \frac{k}{M_{5}} \ M_{grav} \ , \nonumber
\\ &=& -d_{uA} \ (Y_{u})_{ij} \ M_{grav}
\ + \ (a_{u})_{ij} \ \beta_{i} \alpha_{j} \ \frac{k}{M_{5}} \ M_{grav} \ , 
\\ A_{dij} &=& -d_{dA} \ (Y_{d})_{ij} \ M_{grav} 
\ + \ (a_{d})_{ij} \ \gamma_{i} \alpha_{j} \ \frac{k}{M_{5}} \ M_{grav} \ ,
\\ A_{eij} &=& -d_{dA} \ (Y_{e})_{ij} \ M_{grav}
\ + \ (a_{e})_{ij} \ \epsilon_{i} \delta_{j} \ \frac{k}{M_{5}} \ M_{grav} \ .
\end{eqnarray}
Assuming that the components of 5D couplings $d_{*A}$, $(a_{*})_{ij}$ are $O(1)$,
 we obtain the following estimates on the magnitudes at the scale $M_{5}e^{-kR\pi}$:
\begin{eqnarray}
A_{uij} &\sim& (Y_{u})_{ij} \ M_{grav}
\ + \ \beta_{i} \alpha_{j} \ M_{grav} \ , 
\\ A_{dij} &\sim& (Y_{d})_{ij} \ M_{grav} 
\ + \ \gamma_{i} \alpha_{j} \ M_{grav} \ ,
\\ A_{eij} &\sim& (Y_{e})_{ij} \ M_{grav}
\ + \ \epsilon_{i} \delta_{j} \ M_{grav} \ .
\end{eqnarray}
\\

\subsection{RG Contributions}

\ \ \ Let us estimate the magnitudes of the flavor-violating soft SUSY breaking terms that arise from 
the RG equations involving the Yukawa couplings.
In doing so, we take the specific flavor basis where $Y_{u}$ or $Y_{d}$ and $Y_{e}$ are diagonal.

We first study how $Y_{u}$, $Y_{d}$, $Y_{e}$-diagonal bases change through RG evolutions.
Define the following unitary matrices $U_{*}$:
\begin{eqnarray*}
U_{U} Y_{u} U_{Qu} &=& (diag.) \ , \\
U_{D} Y_{d} U_{Qd} &=& (diag.) \ , \\
U_{E} Y_{e} U_{L} &=& (diag.) \ .
\end{eqnarray*}
Note that $U_{*}$s depend on the renormalization scale because the Yukawa matrices receive RG corrections.
We will calculate how $U_{*}$s vary through RG evolutions.
We have the RG equation below:
\begin{eqnarray}
\mu \frac{{\rm d}}{{\rm d} \mu} (U_{U} Y_{u} U_{Qu}) &=&
(\mu \frac{{\rm d}}{{\rm d} \mu} U_{U}) U_{U}^{\dagger} (U_{U} Y_{u} U_{Qu}) \ + \ 
U_{U} (\mu \frac{{\rm d}}{{\rm d} \mu} Y_{u}) U_{Qu} \ + \
(U_{U} Y_{u} U_{Qu}) U_{Qu}^{\dagger} (\mu \frac{{\rm d}}{{\rm d} \mu} U_{Qu}) \nonumber \\
&=& (\mu \frac{{\rm d}}{{\rm d} \mu} U_{U}) U_{U}^{\dagger} (U_{U} Y_{u} U_{Qu}) \nonumber \\
&+& \frac{1}{16 \pi^{2}} \ U_{U} \ \{ \ Y_{u} Y_{d}^{\dagger} Y_{d} + 3 Y_{u} Y_{u}^{\dagger} Y_{u} + 
3 {\rm tr}[Y_{u}^{\dagger}Y_{u}] Y_{u} + {\rm tr}[Y_{D}^{\dagger}Y_{D}] Y_{u} \nonumber \\
& & - \ (\frac{13}{15} g_{1}^{2} + 3 g_{2}^{2} + \frac{16}{3} g_{3}^{2})Y_{u} \ \} \ U_{Qu} \nonumber \\
&+& (U_{U} Y_{u} U_{Qu}) U_{Qu}^{\dagger} (\mu \frac{{\rm d}}{{\rm d} \mu} U_{Qu}) \ ,
\end{eqnarray}
 where $Y_{D}$ is neutrino Dirac coupling which appears 
 if there exist singlet neutrinos lighter than 
 the KK scale, from which the RG equations are calculated.
We hereafter adopt GUT normalization for $g_{1}$.
From (48), we see that, to keep $U_{U}Y_{u}U_{Qu}$ diagonal through RG evolutions, 
 it is sufficient to have
\begin{eqnarray}
\mu \frac{{\rm d}}{{\rm d} \mu} U_{U} &=& 0 \ , \\
\mu \frac{{\rm d}}{{\rm d} \mu} U_{Qu} &=& -\frac{1}{16 \pi^{2}} \ 
({\rm off-diagonal \ components \ of} \ Y_{d}^{\dagger} Y_{d}) \ U_{Qu} \ .
\end{eqnarray}
In a similar manner, we obtain the following sufficient conditions for other $U_{*}$s:
\begin{eqnarray}
\mu \frac{{\rm d}}{{\rm d} \mu} U_{D} &=& 0 \ , \\
\mu \frac{{\rm d}}{{\rm d} \mu} U_{Qd} &=& -\frac{1}{16 \pi^{2}} \ 
({\rm off-diagonal \ components \ of} \ Y_{u}^{\dagger} Y_{u}) \ U_{Qd} \ , \\
\mu \frac{{\rm d}}{{\rm d} \mu} U_{E} &=& 0 \ , \\
\mu \frac{{\rm d}}{{\rm d} \mu} U_{L} &=& -\frac{1}{16 \pi^{2}} \ 
({\rm off-diagonal \ components \ of} \ Y_{D}^{\dagger} Y_{D}) \ U_{L} \ .
\end{eqnarray}
\\

Now that we know how $Y_{u}$, $Y_{d}$, $Y_{e}$-diagonal bases change through RG evolutions,
 we estimate the RG contributions to the A-terms in these bases.
Below is the list of the MSSM RG equations for the A-terms:
\begin{eqnarray}
16 \pi^{2} \mu \frac{{\rm d}}{{\rm d}\mu} (U_{U} A_{u} U_{Qu}) &=& 
3 U_{U} A_{u} Y_{u}^{\dagger} Y_{u} U_{Qu} + 3 U_{U} Y_{u} Y_{u}^{\dagger} A_{u} U_{Qu} \nonumber \\
&+& (U_{U} A_{u} U_{Qu}) ({\rm diagonal \ part \ of} \ U_{Qu}^{\dagger} Y_{d}^{\dagger} Y_{d} U_{Qu}) + 2 U_{U} Y_{u} Y_{d}^{\dagger} A_{d} U_{Qu} \nonumber \\
&+& 2 ( \ 3 {\rm tr}[ Y_{u}^{\dagger} A_{u} ] 
- \frac{13}{15} g_{1}^{2} M^{a=1}_{1/2} - 3 g_{2}^{2} M^{a=2}_{1/2} - \frac{16}{3} g_{3}^{2} M^{a=3}_{1/2} \ ) (U_{U} Y_{u} U_{Qu}) \nonumber \\
&+& ( \ 3 {\rm tr}[ Y_{u}^{\dagger} Y_{u} ] - \frac{13}{15} g_{1}^{2} - 3 g_{2}^{2} - \frac{16}{3} g_{3}^{2} \ ) (U_{U} A_{u} U_{Qu}) \nonumber \\
&+& {\rm tr}[ Y_{D}^{\dagger} Y_{D} ] (U_{U} A_{u} U_{Qu}) + {\rm tr}[ Y_{D}^{\dagger} A_{D} ] (U_{U} Y_{u} U_{Qu}) \ ,
\\
16 \pi^{2} \mu \frac{{\rm d}}{{\rm d}\mu} (U_{D} A_{d} U_{Qd}) &=& 
3 U_{D} A_{d} Y_{d}^{\dagger} Y_{d} U_{Qd} + 3 U_{D} Y_{d} Y_{d}^{\dagger} A_{d} U_{Qd} \nonumber \\
&+& (U_{D} A_{d} U_{Qd}) ({\rm diagonal \ part \ of} \ U_{Qd}^{\dagger} Y_{u}^{\dagger} Y_{u} U_{Qd}) + 2 U_{D} Y_{d} Y_{u}^{\dagger} A_{u} U_{Qd} \nonumber \\
&+& 2 ( \ 3 {\rm tr}[ Y_{d}^{\dagger} A_{d} ] + {\rm tr}[ Y_{e}^{\dagger} A_{e} ]
- \frac{7}{15} g_{1}^{2} M^{a=1}_{1/2} - 3 g_{2}^{2} M^{a=2}_{1/2} - \frac{16}{3} g_{3}^{2} M^{a=3}_{1/2} \ ) (U_{D} Y_{d} U_{Qd})\nonumber \\
&+& ( \ 3 {\rm tr}[ Y_{d}^{\dagger} Y_{d} ] + {\rm tr}[ Y_{e}^{\dagger} Y_{e} ] 
- \frac{7}{15} g_{1}^{2} - 3 g_{2}^{2} - \frac{16}{3} g_{3}^{2} \ ) (U_{D} A_{d} U_{Qd}) \ ,
\\
16 \pi^{2} \mu \frac{{\rm d}}{{\rm d}\mu} (U_{E} A_{e} U_{L}) &=& 
3 U_{E} A_{e} Y_{e}^{\dagger} Y_{e} U_{L} + 3 U_{E} Y_{e} Y_{e}^{\dagger} A_{e} U_{L} \nonumber \\
&+& 2 ( \ 3 {\rm tr}[ Y_{d}^{\dagger} A_{d} ] + {\rm tr}[ Y_{e}^{\dagger} A_{e} ]
- \frac{9}{5} g_{1}^{2} M^{a=1}_{1/2} - 3 g_{2}^{2} M^{a=2}_{1/2} \ ) (U_{E} Y_{e} U_{L}) \nonumber \\
&+& ( \ 3 {\rm tr}[ Y_{d}^{\dagger} Y_{d} ] + {\rm tr}[ Y_{e}^{\dagger} Y_{e} ] 
- \frac{9}{5} g_{1}^{2} - 3 g_{2}^{2} \ ) (U_{E} A_{e} U_{L})\nonumber \\
&+& (U_{E} A_{e} U_{L}) ({\rm diagonal \ part \ of} \ U_{L}^{\dagger} Y_{D}^{\dagger} Y_{D} U_{L}) + 2 U_{E} Y_{e} Y_{D}^{\dagger} A_{D} U_{L} \ ,
\end{eqnarray}
 where $Y_{D}$ and $A_{D}$ respectively indicate neutrino Dirac coupling and its corresponding A-term.

Note that the magnitudes of the components of the Yukawa couplings in each basis are given by
 ($\delta_{ij}$ is the ordinary Kronecker's delta):
\begin{eqnarray*}
(U_{U} Y_{u} U_{Qu})_{ij} &\sim& \beta_{i}\alpha_{i} \ \delta_{ij} \ , \ \ \ \ \ 
(U_{U} Y_{d} U_{Qu})_{ij} \ \sim \ \gamma_{i}\alpha_{j} \ , \\
(U_{D} Y_{u} U_{Qd})_{ij} &\sim& \beta_{i}\alpha_{j} \ , \ \ \ \ \ 
(U_{D} Y_{d} U_{Qd})_{ij} \ \sim \ \gamma_{i}\alpha_{i} \ \delta_{ij} \ , \\
(U_{E} Y_{e} U_{L})_{ij} &\sim& \epsilon_{i}\delta_{i} \ \delta_{ij} \ , \ \ \ \ \
(U_{E} Y_{D} U_{L})_{ij} \ \sim \ \zeta_{i}\delta_{j} \ ,
\end{eqnarray*}
 where $\zeta_{i}$s indicate the geometrical factors for singlet neutrinos and satisfy $\zeta_{i} \leq 1$.
Note also that the A-terms receive RG corrections which are proportional 
 to the corresponding Yukawa couplings and to the gaugino masses.
We write these terms by $M_{u}, M_{d}, M_{e}, M_{D}$ respectively for $A_{u}, A_{d}, A_{e}, A_{D}$.
They depend on both $M_{grav}$ and $M_{gauge}$.

With these ingredients, we estimate the RG contributions to those parts of A-terms which are not proportional to 
 the corresponding Yukawa couplings, 
 or equivalently the off-diagonal components of $(U_{U} A_{u} U_{Qu})$, $(U_{D} A_{d} U_{Qd})$, $(U_{E} A_{e} U_{L})$.
In the right hand sides of (55-57), the second lines determine the magnitudes of the RG contributions.
We thus obtain the following estimates ($i \neq j$):
\begin{eqnarray}
\Delta (U_{U} A_{u} U_{Qu})_{ij} &\sim& 2 \ \beta_{i} (\alpha_{i})^{2} (\gamma_{3})^{2} \alpha_{j} \ \times \ \frac{1}{16 \pi^{2}} \ \int {\rm d}(\ln \mu) \ M_{d} \ ,
\\
\Delta (U_{D} A_{d} U_{Qd})_{ij} &\sim& 2 \ \gamma_{i} (\alpha_{i})^{2} (\beta_{3})^{2} \alpha_{j} \ \times \ \frac{1}{16 \pi^{2}} \ \int {\rm d}(\ln \mu) \ M_{u} \ ,
\\
\Delta (U_{E} A_{e} U_{L})_{ij} &\sim& 2 \ \epsilon_{i} (\delta_{i})^{2} (\zeta_{3})^{2} \delta_{j} \ \times \ \frac{1}{16 \pi^{2}} \ \int {\rm d}(\ln \mu) \ M_{D} \ .
\end{eqnarray}
Since $M_{u}, M_{d}, M_{D}$ depend on $M_{grav}$ and $M_{gauge}$, so do the magnitudes of the RG contributions above.
\\

Let us estimate the RG contributions to soft SUSY breaking matter mass terms.
Below are those parts of the MSSM RG equations that give rise to 
 flavor-violating soft SUSY breaking mass terms:
\begin{eqnarray}
16 \pi^{2} \mu \frac{{\rm d}}{{\rm d}\mu} (U_{Qu}^{\dagger} m_{Q}^{2} U_{Qu}) &\supset& 
U_{Qu}^{\dagger} Y_{u}^{\dagger} Y_{u} m_{Q}^{2} U_{Qu} + U_{Qu}^{\dagger} m_{Q}^{2} Y_{u}^{\dagger} Y_{u} U_{Qu} \nonumber \\
&+& 2 U_{Qu}^{\dagger} Y_{u}^{\dagger} m_{U}^{2} Y_{u} U_{Qu} + 2 (U_{Qu}^{\dagger} Y_{u}^{\dagger} Y_{u} U_{Qu}) m_{H_{u}}^{2} \nonumber \\
&+& ({\rm diagonal \ parts \ of} \ U_{Qu}^{\dagger} Y_{d}^{\dagger} Y_{d} U_{Qu})  (U_{Qu}^{\dagger} m_{Q}^{2} U_{Qu}) \nonumber \\
&+& (U_{Qu}^{\dagger} m_{Q}^{2} U_{Qu}) ({\rm diagonal \ parts \ of} \ U_{Qu}^{\dagger} Y_{d}^{\dagger} Y_{d} U_{Qu}) \nonumber \\
&+& 2 U_{Qu}^{\dagger} Y_{d}^{\dagger} m_{D}^{2} Y_{d} U_{Qu} + 2 (U_{Qu}^{\dagger} Y_{d}^{\dagger} Y_{d} U_{Qu}) m_{H_{d}}^{2} \nonumber \\
&+& 2 U_{Qu}^{\dagger}A_{u}^{\dagger}A_{u}U_{Qu} + 2 U_{Qu}^{\dagger}A_{d}^{\dagger}A_{d}U_{Qu} \ ,
\\
16 \pi^{2} \mu \frac{{\rm d}}{{\rm d}\mu} (U_{U} m_{U}^{2} U_{U}^{\dagger}) &\supset& 
2 U_{U} Y_{u} Y_{u}^{\dagger} m_{U}^{2} U_{U}^{\dagger} + 2 U_{U} m_{U}^{2} Y_{u} Y_{u}^{\dagger} U_{U}^{\dagger} \nonumber \\
&+& 4 U_{U} Y_{u} m_{Q}^{2} Y_{u}^{\dagger} U_{U}^{\dagger} + 4 (U_{U} Y_{u} Y_{u}^{\dagger} U_{U}^{\dagger}) m_{H_{u}}^{2} \nonumber \\
&+& 4 U_{U}A_{u}A_{u}^{\dagger}U_{U}^{\dagger} \ ,
\\ 
16 \pi^{2} \mu \frac{{\rm d}}{{\rm d}\mu} (U_{D} m_{D}^{2} U_{D}^{\dagger}) &\supset& 
2 U_{D} Y_{d} Y_{d}^{\dagger} m_{D}^{2} U_{D}^{\dagger} + 2 U_{D} m_{D}^{2} Y_{d} Y_{d}^{\dagger} U_{D}^{\dagger} \nonumber \\
&+& 4 U_{D} Y_{d} m_{Q}^{2} Y_{d}^{\dagger} U_{D}^{\dagger} + 4 (U_{D} Y_{d} Y_{d}^{\dagger} U_{D}^{\dagger}) m_{H_{d}}^{2} \nonumber \\
&+& 4 U_{D}A_{d}A_{d}^{\dagger}U_{D}^{\dagger} \ ,
\\ 16 \pi^{2} \mu \frac{{\rm d}}{{\rm d}\mu} (U_{L}^{\dagger} m_{L}^{2} U_{L}) &\supset& 
U_{L}^{\dagger} Y_{e}^{\dagger} Y_{e} m_{L}^{2} U_{L} + U_{L}^{\dagger} m_{L}^{2} Y_{e}^{\dagger} Y_{e} U_{L} \nonumber \\
&+& 2 U_{L}^{\dagger} Y_{e}^{\dagger} m_{E}^{2} Y_{e} U_{L} + 2 (U_{L}^{\dagger} Y_{e}^{\dagger} Y_{e} U_{L}) m_{H_{d}}^{2} \nonumber \\
&+& 2 U_{L}^{\dagger}A_{e}^{\dagger}A_{e}U_{L} \nonumber \\
&+& ({\rm diagonal \ parts \ of} \ U_{L}^{\dagger} Y_{D}^{\dagger} Y_{D} U_{L}) (U_{L}^{\dagger} m_{L}^{2} U_{L}) \nonumber \\
&+& (U_{L}^{\dagger} m_{L}^{2} U_{L}) ({\rm diagonal \ parts \ of} \ U_{L}^{\dagger} Y_{D}^{\dagger} Y_{D} U_{L}) \nonumber \\
&+& 2 U_{L}^{\dagger} Y_{D}^{\dagger} m_{N}^{2} Y_{D} U_{L}
+ 2 (U_{L}^{\dagger} Y_{D}^{\dagger} Y_{D} U_{L}) m_{H_{u}}^{2} + 2 U_{L}^{\dagger}A_{D}^{\dagger}A_{D}U_{L} \ ,
\\
16 \pi^{2} \mu \frac{{\rm d}}{{\rm d}\mu} (U_{E} m_{E}^{2} U_{E}^{\dagger}) &\supset& 
2 U_{E} Y_{e} Y_{e}^{\dagger} m_{E}^{2} U_{E}^{\dagger} + 2 U_{E} m_{E}^{2} Y_{e} Y_{e}^{\dagger} U_{E}^{\dagger} \nonumber \\
&+& 4 U_{E} Y_{e} m_{L}^{2} Y_{e}^{\dagger} U_{E}^{\dagger} + 4 (U_{E} Y_{e} Y_{e}^{\dagger} U_{E}^{\dagger}) m_{H_{d}}^{2} \nonumber \\
&+& 4 U_{E}A_{e}A_{e}^{\dagger}U_{E}^{\dagger} \ .
\end{eqnarray}

We first focus on the differences among the diagonal components of different flavors.
From (61), the difference between the components 
 $(U_{Qu}^{\dagger} m_{Q}^{2} U_{Qu})_{ii}$ and $(U_{Qu}^{\dagger} m_{Q}^{2} U_{Qu})_{jj}$
 that arises through RG evolutions is given by ($i>j$) :
\begin{eqnarray}
\Delta \{ \ (U_{Qu}^{\dagger} m_{Q}^{2} U_{Qu})_{ii} - (U_{Qu}^{\dagger} m_{Q}^{2} U_{Qu})_{jj} \ \} 
&\sim& 2 (\alpha_{i})^{2}(\beta_{i})^{2} \ \frac{1}{16 \pi^{2}} \int {\rm d}(\ln \mu) \ 
( \ m_{Q}^{2} + m_{U}^{2} + m_{H_{u}}^{2} + M_{u}^{2} \ ) \nonumber
\\
&+& 2 (\alpha_{i})^{2}(\gamma_{3})^{2} \ \frac{1}{16 \pi^{2}} \int {\rm d}(\ln \mu) \ 
( \ m_{Q}^{2} + m_{D}^{2} + m_{H_{d}}^{2} + M_{d}^{2} \ ) \ , \nonumber \\
\end{eqnarray}
 where we neglected the terms proportional to $(\alpha_{j})^{2}$ because 
 they are smaller than those proportional to $(\alpha_{i})^{2}$.
Similarly, the difference between $(U_{L}^{\dagger} m_{L}^{2} U_{L})_{ii}$ and $(U_{L}^{\dagger} m_{L}^{2} U_{L})_{jj}$
 is given, from (64), by
\begin{eqnarray}
\Delta \{ \ (U_{L}^{\dagger} m_{L}^{2} U_{L})_{ii} - (U_{L}^{\dagger} m_{L}^{2} U_{L})_{jj} \ \} 
&\sim& 2 \ (\delta_{i})^{2}(\epsilon_{i})^{2} \ \frac{1}{16 \pi^{2}} \int {\rm d}(\ln \mu) \ 
( \ m_{L}^{2} + m_{E}^{2} + m_{H_{d}}^{2} + M_{e}^{2} \ ) \nonumber
\\
&+& 2 \ (\delta_{i})^{2}(\zeta_{3})^{2} \ \frac{1}{16 \pi^{2}} \int {\rm d}(\ln \mu) \ 
( \ m_{L}^{2} + m_{N}^{2} + m_{H_{u}}^{2} + M_{D}^{2} \ ) \ . \nonumber \\
\end{eqnarray}
On the other hand, the differences among the diagonal components of SU(2) singlet soft mass terms
 follow different formulae.
From (62, 63, 65), we have
\begin{eqnarray}
\Delta \{ \ (U_{U}^{\dagger} m_{U}^{2} U_{U})_{ii} - (U_{U}^{\dagger} m_{U}^{2} U_{U})_{jj} \ \} 
&\sim& 4 \ (\beta_{i})^{2}(\alpha_{i})^{2} \ \frac{1}{16 \pi^{2}} \int {\rm d}(\ln \mu) \ 
( \ m_{U}^{2} + m_{Q}^{2} + m_{H_{u}}^{2} + M_{u}^{2} \ ) \ , \nonumber \\
\\
\Delta \{ \ (U_{D}^{\dagger} m_{D}^{2} U_{D})_{ii} - (U_{D}^{\dagger} m_{D}^{2} U_{D})_{jj} \ \} 
&\sim& 4 \ (\gamma_{i})^{2}(\alpha_{i})^{2} \ \frac{1}{16 \pi^{2}} \int {\rm d}(\ln \mu) \ 
( \ m_{D}^{2} + m_{Q}^{2} + m_{H_{d}}^{2} + M_{d}^{2} \ ) \ , \nonumber \\
\\
\Delta \{ \ (U_{E}^{\dagger} m_{E}^{2} U_{E})_{ii} - (U_{E}^{\dagger} m_{E}^{2} U_{E})_{jj} \ \} 
&\sim& 4 \ (\epsilon_{i})^{2}(\delta_{i})^{2} \ \frac{1}{16 \pi^{2}} \int {\rm d}(\ln \mu) \ 
( \ m_{E}^{2} + m_{L}^{2} + m_{H_{d}}^{2} + M_{e}^{2} \ ) \ . \nonumber \\
\end{eqnarray}

We next study the off-diagonal components.
In (61), terms $2 U_{Qu}^{\dagger} Y_{d}^{\dagger} m_{D}^{2} Y_{d} U_{Qu}$,
 $2 (U_{Qu}^{\dagger} Y_{d}^{\dagger} Y_{d} U_{Qu}) m_{H_{d}}^{2}$, 
 $2 U_{Qu}^{\dagger}A_{d}^{\dagger}A_{d}U_{Qu}$ generate off-diagonal components, whose magnitudes are given by
 ($i \neq j$)
\begin{eqnarray}
\Delta (U_{Qu}^{\dagger} m_{Q}^{2} U_{Qu})_{ij}
&\sim& 2 \ \alpha_{i}(\gamma_{3})^{2}\alpha_{j} \ \frac{1}{16 \pi^{2}} \int {\rm d}(\ln \mu) \ 
( \ m_{D}^{2} + m_{H_{d}}^{2} + M_{d}^{2} \ ) \ .
\end{eqnarray}
Similarly, we have
\begin{eqnarray}
\Delta (U_{L}^{\dagger} m_{L}^{2} U_{L})_{ij}
&\sim& 2 \ \delta_{i}(\zeta_{3})^{2}\delta_{j} \ \frac{1}{16 \pi^{2}} \int {\rm d}(\ln \mu) \ 
( \ m_{N}^{2} + m_{H_{u}}^{2} + M_{D}^{2} \ ) \ .
\end{eqnarray}
On the other hand, RG contributions to the off-diagonal components of 
 $(U_{U}^{\dagger} m_{U}^{2} U_{U})$ arise from those of $(U_{Q_{u}}^{\dagger} m_{Q}^{2} U_{Q_{u}})$
 and $(U_{U} A_{u} U_{Qu})$ 
 via terms $4 U_{U} Y_{u} m_{Q}^{2} Y_{u}^{\dagger} U_{U}^{\dagger}$, 
 $4 U_{U} A_{u} A_{u}^{\dagger} U_{U}^{\dagger}$ in (62).
From (58), (71), (62), we obtain the following estimate on 
 the magnitudes of the off-diagonal components ($i \neq j$):
\begin{eqnarray}
\Delta (U_{U}^{\dagger} m_{U}^{2} U_{U})_{ij}
&\sim& 8 \ \beta_{i}(\alpha_{i})^{2}(\gamma_{3})^{2}(\alpha_{j})^{2}\beta_{j}
 \ \left( \frac{1}{16 \pi^{2}} \right)^{2} \
 \int {\rm d}(\ln \mu) \ \int {\rm d}(\ln \mu^{\prime}) \ 
( \ m_{D}^{2} + m_{H_{d}}^{2} + M_{d}^{2} \ ) \nonumber
\\
&+& 16 \ \beta_{i}(\alpha_{i})^{2}(\gamma_{3})^{2}(\alpha_{j})^{2}\beta_{j}
 \ \left( \frac{1}{16 \pi^{2}} \right)^{2} \ 
 \int {\rm d}(\ln \mu) \ \left( M_{u} \int {\rm d}(\ln \mu^{\prime}) M_{d} \right) \ .
\end{eqnarray}
In the same way, we obtain the following estimates on the RG contributions to
 the off-diagonal components of $(U_{D}^{\dagger} m_{D}^{2} U_{D})$, $(U_{E}^{\dagger} m_{E}^{2} U_{E})$:
\begin{eqnarray}
\Delta (U_{D}^{\dagger} m_{D}^{2} U_{D})_{ij} 
&\sim& 8 \ \gamma_{i}(\alpha_{i})^{2}(\beta_{3})^{2}(\alpha_{j})^{2}\gamma_{j}
 \ \left( \frac{1}{16 \pi^{2}} \right)^{2} \
 \int {\rm d}(\ln \mu) \ \int {\rm d}(\ln \mu^{\prime}) \ 
( \ m_{U}^{2} + m_{H_{u}}^{2} + M_{u}^{2} \ ) \nonumber
\\
&+& 16 \ \gamma_{i}(\alpha_{i})^{2}(\beta_{3})^{2}(\alpha_{j})^{2}\gamma_{j}
 \ \left( \frac{1}{16 \pi^{2}} \right)^{2} \ 
 \int {\rm d}(\ln \mu) \ \left( M_{d} \int {\rm d}(\ln \mu^{\prime}) M_{u} \right) \ ,
\\ \nonumber
\\
\Delta (U_{E}^{\dagger} m_{E}^{2} U_{E})_{ij}
&\sim& 8 \ \epsilon_{i}(\delta_{i})^{2}(\zeta_{3})^{2}(\delta_{j})^{2}\epsilon_{j}
 \ \left( \frac{1}{16 \pi^{2}} \right)^{2} \
 \int {\rm d}(\ln \mu) \ \int {\rm d}(\ln \mu^{\prime}) \ 
( \ m_{N}^{2} + m_{H_{u}}^{2} + M_{D}^{2} \ ) \nonumber
\\
&+& 16 \ \epsilon_{i}(\delta_{i})^{2}(\zeta_{3})^{2}(\delta_{j})^{2}\epsilon_{j}
 \ \left( \frac{1}{16 \pi^{2}} \right)^{2} \ 
 \int {\rm d}(\ln \mu) \ \left( M_{e} \int {\rm d}(\ln \mu^{\prime}) M_{D} \right) \ .
\end{eqnarray}
\\
\\

Finally, we briefly discuss whether this model gives a realistic mass spectrum consistent with 
 the bounds on flavor-violating processes.

For cases without messenger fields, 
 i.e. when gaugino mediation is the only source of soft SUSY breaking masses,
 the paper \cite{gaugino mediation + bulk matter} showed that there exist mass spectra below TeV scale
 that satisfy all experimental bounds.
However, $\sim 0.1$ suppression on the term $(A_{e})_{21}$ relative to its natural scale ($\sim \epsilon_{2}\delta_{1}M_{grav}$)
 is required to evade the bound on $\mu \rightarrow e \gamma$ branching ratio.
Other soft SUSY breaking terms are less constrained.

If there are 1 to several messenger pairs, the resultant mass spectra are more likely to evade the experimental bounds
 because gauge mediation contributes solely to flavor-universal soft SUSY breaking terms.
\\

\section{Signatures of the Model}

\ \ \ In the previous section, we saw that
 the bulk matter RS model combined with 5D MSSM predicts 
 a unique flavor structure of gravity mediation contributions to flavor-violating soft terms.
We here discuss the ways to observe this structure
 through future collider experiments.

Focus on the flavor compositions of SUSY matter particle mass eigenstates.
Due to flavor-violating soft mass terms $(m^{2}_{*})_{ij}$ 
 and flavor-violating A-terms $(A_{*})_{ij}$,
 SUSY particles of different flavors mix in one mass eigenstate,
 whose flavor composition reflects the relative size of the flavor-violating terms.
Since sparticles of different flavors decay into different SM particles 
 (plus the lightest or the next-to-lightest SUSY particle),
 one can measure the flavor composition
 by detecting the decay products of that mass eigenstate,
 counting the event numbers of different decay modes
 and calculating their ratios.
These ratios are connected to the structure of flavor-violating soft SUSY breaking terms
 and make it possible to experimentally test the predictions of the bulk matter RS model.

Below we formulate the relation between flavor-violating terms and sparticle flavor mixings.
In the first subsection, we interpret the predictions of the bulk matter RS model 
 in terms of the flavor mixing ratios of sparticle mass eigenstates.
In the next subsection, we look into the predictions of models other than the bulk matter RS model
 and discuss whether or not it is possible to distinguish different models.
\\

\ \ \ Consider the situation where sparticle ``a" with soft SUSY breaking mass $m_{a}^{2}$
 mixes with sparticle ``b" with soft mass $m_{b}^{2}$
 through mixing term $\Delta m^{2}$. 
The mass matrix in the basis of $(a, b)$ is given by
\begin{eqnarray*}
\left(
\begin{array}{cc}
m_{a}^{2} & \Delta m^{2} \\
\Delta m^{2} & m_{b}^{2}
\end{array}
\right) \ .
\end{eqnarray*}
The mass eigenstates are derived by diagonalizing the matrix above.
If $\vert m_{a}^{2} - m_{b}^{2} \vert >> 2 \vert \Delta m^{2} \vert$ holds,
 the mixing ratios of ``a" and ``b" in the two mass eigenstates are approximately given by
\begin{eqnarray*}
\vert m_{a}^{2} - m_{b}^{2} \vert \ : \ \vert \Delta m^{2} \vert \ , \ \ \ \ \ \ \vert \Delta m^{2} \vert \ : \ \vert m_{a}^{2} - m_{b}^{2} \vert \ .
\end{eqnarray*}
\\

\subsection{Predictions of the Bulk Matter RS Model}

\ \ \ The bulk matter RS model
 predicts a nontrivial structure of flavor-violating soft SUSY breaking terms, 
 given by (41, 45-47, 66-75).
This structure can be translated into the flavor composition of each SUSY particle mass eigenstate.
One subtlety is that the flavor-violating terms contain two different SUSY breaking mass scales,
 namely, the IR-scale-suppressed gravity mediation scale, $M_{grav}$, 
 and the gauge mediation scale, $M_{gauge}$;
 flavor-violating gravity mediation contributions depend solely on $M_{grav}$,
 whereas RG contributions are proportional to the net soft SUSY breaking mass scale
 that depends both on $M_{grav}$ and $M_{gauge}$.
The relative size of these scales affects the predictions on the flavor compositions.
We consider three cases with $M_{grav} \gtrsim M_{gauge}$, $M_{grav} < M_{gauge}$ and $M_{grav} << M_{gauge}$,
 whose precise definitions will be given each time.
These cases lead to different predictions.
\\

\subsubsection{Case with $M_{grav} \gtrsim M_{gauge}$}

\ \ \ In this case, flavor-universal soft SUSY breaking masses, $m^{2}_{*}$,
 and gaugino masses, $M_{**}$, 
 are of the same magnitude as the gravity mediation scale, $M_{grav}$.
The differences among the diagonal components of different flavors
 come from the gravity mediation contributions (41) and the RG contributions (66-70).
Since we now have $m^{2}_{*} \sim M_{grav}^{2}, \ M_{**} \sim M_{grav}$,
 terms (41) surpass terms (66-70).
Hence we may make the following approximations for $i > j$ in any flavor basis:
\begin{eqnarray}
(m_{Q}^{2})_{ii} \ - \ (m_{Q}^{2})_{jj} &\sim& \alpha_{i}^{2} \ M_{grav}^{2}
\end{eqnarray}
 and similar formulae with $(U, \beta), \ (D, \gamma), \ (L, \delta), \ (E, \epsilon)$
 relacing $(Q, \ \alpha)$ in the above formula.

In a similar manner, in any flavor basis,
 the A-terms are approximated by
\begin{eqnarray}
(A_{u})_{ij} &\supset& \beta_{i}\alpha_{j} \ M_{grav} \ , \ \ \ \ \ 
(A_{d})_{ij} \ \supset \ \gamma_{i}\alpha_{j} \ M_{grav} \ , \ \ \ \ \ 
(A_{e})_{ij} \ \supset \ \epsilon_{i}\delta_{j} \ M_{grav} \ ,
\end{eqnarray}
 and the off-diagonal components of soft SUSY breaking mass terms
 are by ($i \neq j$)
\begin{eqnarray}
(m_{Q}^{2})_{ij} &\sim& \alpha_{i}\alpha_{j} \ M_{grav}^{2}
\end{eqnarray}
 and similar formulae with $(U, \beta), \ (D, \gamma), \ (L, \delta), \ (E, \epsilon)$
 relacing $(Q, \ \alpha)$ in the above formula.

Sparticle $Q_{i}$ mixes with sparticle $Q_{j} \ (j \neq i)$ through the term $(m_{Q}^{2})_{ij}$
 and with $U_{k}$ or $D_{k} \ (k \neq i)$ through the A-terms and the VEVs of the Higgs bosons.
In this way, there appears a mass eigenstate that consists mainly of $Q_{i}$
 and partly of $Q_{j}$ and $U_{k}$ or $D_{k}$,
 which we hereafter call ``almost $Q_{i}$ mass eigenstate".
From (76, 78), the mixing ratio of $Q_{j}$ in ``almost $Q_{i}$" mass eigenstate
 is given by
\begin{eqnarray}
\frac{ \vert (m_{Q}^{2})_{ij} \vert }{ \vert (m_{Q}^{2})_{ii} - (m_{Q}^{2})_{jj} \vert } &\simeq& 
 \frac{ \alpha_{i}\alpha_{j} M_{grav}^{2} }{ (\alpha_{i})^{2} M_{grav}^{2} }
 \ \sim \ \frac{ \alpha_{j} }{ \alpha_{i} }
\end{eqnarray}
 for $i>j$, and by
\begin{eqnarray}
\frac{ \vert (m_{Q}^{2})_{ij} \vert }{ \vert (m_{Q}^{2})_{ii} - (m_{Q}^{2})_{jj} \vert } &\simeq&
 \frac{ \alpha_{i}\alpha_{j} M_{grav}^{2} }{ (\alpha_{j})^{2} M_{grav}^{2} }
 \ \sim \ \frac{ \alpha_{i} }{ \alpha_{j} }
\end{eqnarray}
 for $i<j$.
On the other hand, the mixing ratio of $U_{j}$ in the up-sector of ``almost $Q_{i}$ mass eigenstate"
 is given by ($i \neq j$)
\begin{eqnarray}
\frac{ v_{u} \ \vert (A_{u})_{ji} \vert }{ \vert m_{Q}^{2} - m_{U}^{2} \vert } &\sim&
 \frac{ v_{u} \ \beta_{j}\alpha_{i} M_{grav} }{ M_{susy}^{2} }
 \ \sim \ \beta_{j}\alpha_{i} \ \frac{ v_{u} }{ M_{susy} } \ ,
\end{eqnarray}
 where we used the fact that the difference between the flavor-universal masses of 
 SU(2) doublet and singlet squarks is of the same magnitude as 
 the soft SUSY breaking mass scale itself, denoted by $M_{susy}$.

The mixing ratios in other mass eigenstates follow similar formulae.
There is a subtlety about the ratio of $L_{j}$ in ``almost $L_{i}$ mass eigenstate"
 because we have $3 \delta_{1} \sim \delta_{2} \sim \delta_{3}$ and
 the approximation used to derive (79-81) is no longer valid.
Actually, the mixing ratio of $L_{j}$ in ``almost $L_{i}$ mass eigenstate"
 is $O(1)$ for any $i, j$.
\\

\subsubsection{Case with $M_{grav} << M_{gauge}$}

\ \ \ In this sub-subsection, we concentrate on the case where 
 the ratio $M_{grav}/M_{gauge}$ is so small that the RG contributions to 
 flavor-violating soft SUSY breaking terms, (58-60, 66-75),
 are of the same magnitude as or larger than the gravity mediation contributions, (41, 45-47).

In these cases, the mixing ratio of $Q_{j}$ in ``almost $Q_{i}$ mass eigenstate"
 is given, from (66, 71), by ($i > j$)
\begin{eqnarray}
\frac{ \vert (m_{Q}^{2})_{ij} \vert }{ \vert (m_{Q}^{2})_{ii} - (m_{Q}^{2})_{jj} \vert } &\sim& 
 \frac{ \alpha_{i} (\gamma_{3})^{2} \alpha_{j} }{ (\alpha_{i})^{2}(\beta_{i})^{2} + (\alpha_{i})^{2}(\gamma_{3})^{2} }
 \ \sim \ \frac{ \alpha_{j} }{ \alpha_{i} } \ \frac{ (\gamma_{3})^{2} }{ (\beta_{i})^{2} + (\gamma_{3})^{2} } \ ,
\end{eqnarray}
 in the flavor basis where $Y_{u}$ is diagonalized.
Here we used the fact that the integrand of the right hand side of (66)
 and that of (71) are of the same magnitude.
On the other hand, from (58), 
 the mixing ratio of $U_{j}$ in the up-sector of ``almost $Q_{i}$ mass eigenstate" is given by ($i \neq j$)
\begin{eqnarray}
\frac{ v_{u} \vert (A_{u})_{ji} \vert }{ \vert m_{Q}^{2} - m_{U}^{2}\vert } &\sim& 
 2 \beta_{j} (\alpha_{j})^{2} (\gamma_{3})^{2} \alpha_{i} \ \frac{ v_{u} }{ M_{gauge} }
\end{eqnarray}
 in $Y_{u}$-diagonal basis.
Here we approximated the difference of flavor-conserving masses of SU(2) doublet squarks and singlet up-type squarks
 by $M_{gauge}$.
The mixing ratio of $D_{j}$ in the down-sector of ``almost $Q_{i}$ mass eigenstate" in $Y_{d}$-diagonal basis
 takes a similar expression.
The same discussion applies to the mixings in ``almost $L_{i}$ mass eigenstate".

The mixing ratios in ``almost $U_{i}$ mass eigenstate" follow different formulae.
From (68, 73), the ratio of $U_{j}$ is given by ($i > j$)
\begin{eqnarray}
\frac{ \vert (m_{U}^{2})_{ij} \vert }{ \vert (m_{U}^{2})_{ii} - (m_{U}^{2})_{jj} \vert } &\sim& 
 \frac{ 24 \ \beta_{i} (\alpha_{i})^{2} (\gamma_{3})^{2} (\alpha_{j})^{2} \beta_{j} }
 { 4 \ (\beta_{i})^{2} (\alpha_{i})^{2} }
 \ \sim \ 6 \ (\gamma_{3})^{2} (\alpha_{j})^{2} \ \frac{ \beta_{j} }{ \beta_{i} }
\end{eqnarray}
 in $Y_{u}$-diagonal basis.
On the other hand, from (58), the ratio of the up-sector of $Q_{j}$ in ``almost $U_{i}$ mass eigenstate"
 is given by ($i \neq j$)
\begin{eqnarray}
\frac{ v_{u} \vert (A_{u})_{ij} \vert }{ \vert m_{Q}^{2} - m_{U}^{2} \vert } &\sim&
 2 \beta_{i} (\alpha_{i})^{2} (\gamma_{3})^{2} \alpha_{j} \ \frac{ v_{u} }{ M_{gauge} }
\end{eqnarray}
 in $Y_{u}$-diagonal basis.
The same discussion applies to the mixings in ``almost $D_{i}$ mass eigenstate"
 and ``almost $E_{i}$ mass eigenstate".
\\

\subsubsection{Case with $M_{grav} < M_{gauge}$ but not with $M_{grav} << M_{gauge}$}

\ \ \ Consider the case where $M_{grav}$ is slightly smaller than $M_{gauge}$.
Then gravity mediation contributions surpass RG contributions
 for some of the flavor-violating soft SUSY breaking terms,
 and the opposite holds for the other terms.
In these cases, the mixing ratios of sparticle mass eigenstates generally depend on 
 the unknown ratio $M_{grav}/M_{gauge}$ 
 and the model loses its predictive power.

However, certain mixing ratios are more likely to reflect the gravity mediation contributions.
For example, if $M_{grav} \ \gtrsim \ \delta_{3} M_{gauge}$,
 as to terms $(m_{E}^{2})_{ii} - (m_{E}^{2})_{jj}$ and $(m_{E})^{2}_{ij}$,
 the gravity mediation contributions described by (41) are larger than the RG contributions, (70, 75).
Then the mixing ratio of $E_{j}$ in ``almost $E_{i}$ mass eigenstate" is the same as in the case
 with $M_{grav} \gtrsim M_{gauge}$.
Focusing on such mixing ratios, it is still possible to observe the signatures of the model.
\\

\subsection{Comparison with Other Models}

\ \ \ To test the predictions of the bulk matter RS model,
 we must check whether they contain signatures distinguishable from other models.
As an example, we investigate two types of models; 
 one is ``minimal flavor violation", in which RG contributions of the Yukawa couplings
 are the only source of flavor-violating soft SUSY breaking terms.
The other is ``4D gravity mediation", in which gravity mediation
 contributes uniformly to all flavor-violating terms.
We will compare the predictions of these models with the bulk matter RS model
 and discuss the ways to distinguish them.
\\

\subsubsection{Minimal Flavor Violation}

\ \ \ The minimal flavor violation (MFV) scenario leads to the same result as in section 4.1.2,
 i.e. the bulk matter RS model with $M_{grav} << M_{gauge}$.
This is because the argument in section 4.1.2 holds irrespective of gravity mediation contributions.
We thus conclude that it is impossible to experimentally distinguish the bulk matter RS model
 from the MFV scenario
 when we have $M_{grav} << M_{gauge}$, as in 4.1.2.

In contrast, if $M_{grav} \gtrsim M_{gauge}$, 
 the MFV scenario and the bulk matter RS model have distinctively different
 predictions on the mixing ratios in ``almost $U_{i}$, $D_{i}$, $E_{i}$ mass eigenstates" with $i=1,2$.
This is seen by comparing (79-81) ($Q$ relaced by $U, D, E$) with (84, 85);
 the flavor mixings in these mass eigenstates are suppressed at least by 
 $(\alpha_{2})^{2}$ or $(\delta_{2})^{2}$ in ``minimal flavor violation" compared to the bulk matter RS model.
Therefore it is possible to discriminate the two models
 by observing the flavor compositions of 
 ``almost 1st or 2nd generation SU(2) singlet sparticle mass eigenstates".
\\

\subsubsection{4D Gravity Mediation}

\ \ \ We here discuss the case where 4D theory description is valid even at the Planck scale,
 or all matter superfields are confined on the same 4D brane.
Then the gravity mediation contributions are of the same magnitude
 irrespective of flavors.
Of particular interest is the situation where 
 the gravity mediation contributions surpass the flavor-violating RG contributions,
 which is the case when $M_{grav}$ is only slightly smaller than $M_{gauge}$.
In this situation,
 the differences between diagonal components of soft SUSY breaking masses
 $(m_{*}^{2})_{ii} - (m_{*}^{2})_{jj}$, and
 off-diagonal components $(m_{*}^{2})_{ij}$,
 in any flavor basis are of the same magnitude.
Then the mixing ratios in sparticle mass eigenstates are all $O(1)$.
It is easy to distinguish this model from the bulk matter RS model,
 where the mixing ratios of recessive flavors are suppressed by the geometrical factors.
\\

\section{Experimental Studies}

\ \ \ In the previous section, we saw that
 the bulk matter RS model has a unique prediction on the flavor compositions 
 of sparticle mass eigenstates that may be distinguishable from other models.
In this section, we study how to measure the predicted mixing ratios
 through collider experiments.
We focus on the case where $M_{grav} \simeq M_{gauge}$ holds
 or $M_{grav}$ is slightly smaller than $M_{gauge}$,
 and put emphasis on distinguishing
 the bulk matter RS model from the MFV scenario.

The basic strategy is to create a specific mass eigenstate(s) of SUSY matter particles,
 detect its decay products and count the numbers of events of different decay modes.
The branching fractions of different modes reflect the flavor composition of that mass eigenstate.
There are, however, three challenges for this study.

First, we have to detect small flavor components of sparticle mass eigenstates,
 which means that we have to observe \textit{rare} decay events in collider experiments.
For this purpose, the probability of misidentifying the decay products of 
 the dominant mode as those of
 a rare mode must be negligibly small.
For example, the stau components of ``almost smuon mass eigenstates" are detectable
 because SM tau from the stau components,
 when we focus on its hadronic decay, leaves a signal different from muon events.
However, it is impossible to observe the smuon components of ``almost stau mass eigenstates"
 because SM tau from the dominant stau components may decay into SM muon, 
 which mimics the smuon component signal.

Second, we have to extract the decay products of a \textit{specific} mass eigenstate
 in order to compare the data with
 the predictions of the bulk matter RS model.
It is thus required to produce only specific mass eigenstates at a collider.
This is achieved by lepton colliders, such as the ILC \cite{ILC} and the CLIC \cite{CLIC},
 where the center-of-mass energy of a process is fixed.
For example, the flavor-mixing ratios in ``almost SU(2) doublet smuon mass eigenstate"
 and in ``almost SU(2) singlet smuon mass eigenstate" are predicted to be different.
To confirm this prediction, we must produce one of the two eigenstates selectively.
If the latter is lighter than the former,
 we take the center-of-mass energy between their thresholds
 so that only the latter is created on-shell.
We then measure the mixing ratios of the latter eigenstate through its decay products.
In conclusion, lepton colliders are essential when studying
 the flavor compositions of sparticle mass eigenstates.

Finally, we have to focus on ``almost SU(2) singlet mass eigenstates" 
 in order to discriminate the bulk matter RS model from the MFV scenario.
This is understood by comparing the predictions of the bulk matter RS model, (79-81),
 with those of the MFV scenario, (82-85).
Remember that we have
\begin{eqnarray*}
\gamma_{3} &\sim& \tan \beta \ \frac{m_{b}}{v} \ , \ \ \ \ \ \beta_{3} \ \sim \ 1
\end{eqnarray*}
 and we do not know the magnitude of $\zeta_{3}$.
Hence it can be the case that the mixing ratios of $Q_{j}$ 
 in ``almost $Q_{i}$ mass eigenstate", and those of $L_{j}$
 in ``almost $L_{i}$ mass eigenstate" are of the same magnitudes for the bulk matter RS model and 
 the MFV scenario.
In contrast, the mixing ratios of $U_{j}, D_{j}, E_{j}$ in ``almost $U_{i}, D_{i}, E_{i}$ mass eigenstates"
 are of the different magnitudes for the two models
 because the mixing ratios in the MFV scenario, (82-85), are suppressed by the factors
 $(\alpha_{1})^{2}, (\alpha_{2})^{2}, (\delta_{1})^{2}$ or $(\delta_{2})^{2}$
 compared to those in the bulk matter RS model, (79-81).
We further notice that ``almost 3rd generation sparticle mass eigenstates"
 are not suitable for our study
 because the 3rd generation sparticles have significant left-right mixing terms due to their large Yukawa couplings.
We conclude that observing the rare decays of 
 ``almost SU(2) singlet 1st and 2nd generation mass eigenstates"
 is the only way to distinguish the bulk matter RS model and the MFV scenario.

Taking these points into account, we will discuss
 three types of experiments that are feasible at future lepton colliders.
The first type of experiment deals with the rare decay of ``almost SU(2) singlet smuon mass eigenstate" into SM tau,
 which reflects the mixing of singlet smuon with stau.
Another type of experiment deals with the rare decay of ``almost SU(2) singlet smuon mass eigenstate" into SM electron
 or that of ``almost SU(2) singlet selectron" into SM muon,
 which reflects the mixing of singlet smuon and selectron.
The other type of experiment deals with the rare decay of ``almost SU(2) singlet scharm mass eigenstate" into SM top,
 which reflects the mixing of singlet scharm with stop.
For a concrete discussion,
 we make assumptions on the SUSY particle mass spectrum in section 5.1.
We then look into the three types of experiments in section 5.2., 5.3. and 5.4.

``Almost the lighter stau / stop mass eigenstate" are schematically denoted
 by $\tilde{\tau}_{1}$ / $\tilde{t}_{1}$,
 and ``almost singlet selectron / smuon / scharm mass eigenstate" are
 by $\tilde{e}_{R}$ / $\tilde{\mu}_{R}$ / $\tilde{\mu}_{R}$.

It is impossible to do these experiments at hadron colliders.
This is fundamentally because we need to create ``almost SU(2) singlet mass eigenstates" exclusively,
 without creating their ``almost SU(2) doublet" counterparts,
 in order to discriminate the bulk matter RS model from the MFV scenario.
Hadron colliders would necessarily create both eigenstates, and the decay products of the latter
 would contaminate the signals that allow us to distinguish the two scenarios.
It is true that ``almost SU(2) singlet eigenstates" are normally lighter than 
 their ``almost SU(2) doublet" counterparts, and hence the production cross sections
 of the latter are lower even at hadron colliders. 
However, since the two scenarios only predict the orders of magnitudes of
 the branching ratios of rare events, 
 even small contamination from the latter would make it difficult to test the predictions.
\\

\subsection{Assumptions on the Mass Spectrum}

\ \ \ In this subsection, we make plausible assumptions on the SUSY particle mass spectrum
 that are consistent with the bulk matter RS model combined with 5D MSSM.

We assume that squarks are heavier than sleptons and gluino is heavier than Wino and Bino
 because of their SU(3) charges.
Also, Wino is assumed heavier than Bino due to its SU(2) charge.
SU(2) doublet squarks are heavier than singlet squarks, and doublet sleptons are than singlet sleptons.
Since gauge superfields are flat in the bulk, i.e. they have no $y$-dependence,
 they obtain large soft SUSY breaking masses through contact terms on the IR brane.
Therefore gluino tends to be heavier than squarks.
Wino and Bino are heavier than sleptons but lighter than squarks.

The $\mu$-term is assumed larger than Wino and Bino masses, as is normally the case.

We do not specify the mass order among doublet and singlet squarks and sleptons
 because gravity mediation contributions may distort the mass spectrum.
However, we expect that the masses of the 1st and 2nd generation SUSY particles
 are almost degenerate
 because their Yukawa couplings as well as their overlap with the IR brane
 are small.

Gravitino is always the lightest SUSY particle (LSP) because its mass is given by
 \ $\sim $ TeV $\times e^{-kR\pi}$.
The next-to-lightest SUSY particle (NLSP) is
 ``almost SU(2) singlet selectron", ``almost singlet smuon" or ``almost the lighter stau"
 mass eigenstate.
The lifetime of the NLSP satisfies
\begin{eqnarray*}
 c t_{NLSP} &\simeq& 48 \pi 
 \frac{ \vert<F_{\tilde{X}}>\vert^{2}}{m_{NLSP}^5} 
 \simeq  48 \pi \frac{M_{grav}^{2} ( M_{5} e^{-kR\pi} )^{2}}{(m_{NLSP})^5} 
\\
&\simeq& (1.2 \times 10^{-26}) {\rm m} \ \times \ \left( \frac{ M_{grav} }{ {\rm GeV} } \right)^{2} \
\left( \frac{ M_{5}e^{-kR\pi} }{ {\rm GeV} } \right)^{2} \ 
\left( \frac{ 300 {\rm GeV} }{ m_{NLSP} } \right)^{5} \ .
\end{eqnarray*}
We assume that the lifetime is enough long that 
 the NLSP reaches the inner detector before it decays.

The order of the sparticle soft SUSY breaking masses is summarized below:
\begin{eqnarray*}
\tilde{H}_{u}, \tilde{H}_{d} \ > \ \tilde{g} \ > \ \tilde{q}_{L} \ > \ \tilde{q}_{R} 
 \ > \ \chi^{\pm}_{1}, \ \chi^{0}_{2} \ (\fallingdotseq \tilde{W}) \ > \ \chi^{0}_{1} \ (\fallingdotseq \tilde{B})
 \ > \ \tilde{l}_{L} \ > \ \tilde{l}_{R} \ > \ \psi_{3/2} \ .
\end{eqnarray*}
\\

\subsection{Type I - Smuon Rare Decay with Stau-like NLSP}

\ \ \ Consider the case where ``almost the lighter stau mass eigenstate" 
 ($\tilde{\tau}_{1}$) is the NLSP.
Tune the center-of-mass energy of the lepton collider
 between the thresholds of ``almost singlet selectron / smuon mass eigenstates"
 ($\tilde{e}_{R} / \tilde{\mu}_{R}$) and
 ``almost SU(2) doublet selectron  / smuon",
 ``almost the heavier stau" mass eigenstates.
Then $\tilde{e}_{R}, \tilde{\mu}_{R}, \tilde{\tau}_{1}$
 are produced on-shell,
 while other sparticle mass eigenstates are not.

The signal for $\tilde{e}_{R}$ or $\tilde{\mu}_{R}$ pair production 
 is a pair of long-lived charged massive particles, which are NLSP $\tilde{\tau}_{1}$s,
 plus two pairs of hard SM leptons.
Note that, since the masses of $\tilde{e}_{R}$ and $\tilde{\mu}_{R}$ are 
 almost degenerate,
 we cannot detect SM leptons emitted when the heavier one decays into the lighter one.
Normally, we have two SM muons or electrons plus two SM taus in these events
 (we call this ``main mode"), e.g.
\begin{eqnarray}
e e \ \rightarrow \ \tilde{\mu}_{R} \ \tilde{\mu}_{R}
\ \rightarrow \ \mu \ \tau \ \tilde{\tau}_{1} \ \mu \ \tau \ \tilde{\tau}_{1}
\end{eqnarray}
 for the sumon production.
However, due to the small stau components in $\tilde{e}_{R}$ / $\tilde{\mu}_{R}$,
 we may also have one SM muon or electron plus three SM taus in these events
  (we call this ``rare mode"), e.g.
\begin{eqnarray}
e e \ \rightarrow \ \tilde{\mu}_{R} \ \tilde{\mu}_{R}
\ \rightarrow \ \tau \ \tau \ \tilde{\tau}_{1} \ \mu \ \tau \ \tilde{\tau}_{1}
\end{eqnarray}
 for the smuon production.
Requiring hadronic decay of SM taus and taking advantage of the tau vertexing,
 one can reduce the probability of misidentifying a main mode event as a rare mode event
 to a negligible level.

The branching ratio of the rare mode is proportional to the square of the mixing ratio.
The stau component in $\tilde{\mu}_{R}$ plays a dominant role
 because the stau component in $\tilde{e}_{R}$ is much more suppressed.
From (79) with $(Q, \alpha)$ replaced by $(E, \epsilon)$,
 and from (81) with $(Q, U, \alpha, \beta, v_{u})$
 replaced by $(L, E, \delta, \epsilon, v_{d})$,
 the bulk matter RS model predicts that
 the branching ratio of the rare mode is given by
\begin{eqnarray}
Br(\tilde{\mu}_{R} \ \rightarrow \ \tau \ \tau \ \tilde{\tau}_{1}) 
&\sim& \left( \frac{ \epsilon_{2} }{ \epsilon_{3} } \right)^{2}
 \ + \ \left( \epsilon_{2} \delta_{3} \frac{v_{d}}{M_{susy}} \right)^{2} \nonumber
\\
&\sim& \left( \frac{ m_{\mu} }{ m_{\tau} } \right)^{2}
 \ + \ \left( \frac{ m_{\mu} }{ M_{susy} } \right)^{2} \ ,
\end{eqnarray}
 where the first term comes from the mixing with singlet stau
 and the second from the mixing with doublet stau.
If $\tilde{e}_{R}$ is lighter than $\tilde{\mu}_{R}$,
 the branching ratio is reduced by $1/2$ compared to the opposite case
 because a half of $\tilde{\mu}_{R}$s decay into $\tilde{e}_{R}$s.
However, this does not affect the order estimate above.
Since we have $M_{susy} >> m_{\tau}$, the second term is negligible
 and the branching ratio becomes
\begin{eqnarray}
Br(\tilde{\mu}_{R} \ \rightarrow \ \tau \ \tau \ \tilde{\tau}_{1}) &\sim& 0.004 \ .
\end{eqnarray}
Note that the prediction above may change by $O(0.1)-O(10)$
 because we only know the magnitudes of the higher-dimensional couplings
 for soft SUSY breaking terms.

On the other hand, the MFV scenario predicts that
 the branching ratio of the rare mode is given by
\begin{eqnarray}
Br(\tilde{\mu}_{R} \ \rightarrow \ \tau \ \tau \ \tilde{\tau}_{1}) 
 &\sim& \left\{ 6 (\zeta_{3})^{2} (\delta_{2})^{2} \frac{ \epsilon_{2} }{ \epsilon_{3} } \right\}^{2}
 \ + \ \left\{ 2 (\zeta_{3})^{2} (\delta_{2})^{2} \epsilon_{2} \delta_{3} \frac{v_{d}}{M_{susy}} \right\}^{2} \nonumber
\\
&\sim& (\zeta_{3} \delta_{2})^{4} \times 0.1 \ ,
\end{eqnarray}
 where we used (84) and (85) 
 with $(Q, U, \alpha, \beta, \gamma, v_{u})$ replaced by $(L, E, \delta, \epsilon, \zeta, v_{d})$.
Although we cannot determine the magnitude of $\zeta_{3} \delta_{2}$, 
 we expect it to be smaller than $0.1$;
 from (72), we have the following flavor-mixing term for SU(2) doublet smuon and selectron:
\begin{eqnarray*}
(m^{2}_{L})_{12} &\sim& \delta_{1} (\zeta_{3})^{2} \delta_{2} \ M_{susy}^{2}
 \ \sim \ \frac{1}{3} (\zeta_{3} \delta_{2})^{2} \ M_{susy}^{2} \ .
\end{eqnarray*}
For example, 
 with $M_{susy}=500$ GeV, $m_{\tilde{l}_{L}}^{2}=500$ GeV, $M_{\tilde{B}}=M_{\tilde{W}}=750$ GeV,
 $\mu=1000$ GeV, $\tan \beta=10$ and vanshing A-terms,
 taking $\zeta_{3} \delta_{2} = 0.1$ would saturate the current bound on $\mu \rightarrow e \gamma$
 branching ratio, $Br(\mu \rightarrow e \gamma) \leq 1.2 \times 10^{-11}$ \cite{MEGA}.
Hence the branching ratio of the rare mode in the MFV scenario satisfies
\begin{eqnarray}
Br(\tilde{\mu}_{R} \ \rightarrow \ \tau \ \tau \ \tilde{\tau}_{1}) 
&\lesssim& 10^{-5} \ ,
\end{eqnarray}
 which may get smaller if the bound on $Br(\mu \rightarrow e \gamma)$ improves.
We conclude that the branching ratio of the rare mode
 is distinctively smaller in the MFV scenario than in the bulk matter RS model.
\\

\subsection{Type II - NLSP Selectron Rare Decay into Muon, or
 NLSP Smuon Rare Decay into Electron / Tau}

\ \ \ Consider the case where ``almost singlet smuon" or ``almost singlet selectron2 mass eigenstate
 ($\tilde{\mu}_{R}$ or $\tilde{e}_{R}$) is the NLSP and is long-lived.
Tune the center-of-mass energy slightly above the threshold of $\tilde{\mu}_{R}$ / $\tilde{e}_{R}$
 so that they are produced with a low $\beta$ (Lorentz velocity).
Slow long-lived sleptons may be trapped in the inner detector.
According to the paper \cite{NLSP@ILC}, taking $\beta \lesssim 0.2$ is sufficient
 to trap $600$ GeV or lighter sleptons in the inner detector.
We analyze the decay products of these sleptons to study their flavor compositions.
\\

First study the case where $\tilde{e}_{R}$ is lighter than $\tilde{\mu}_{R}$
 and is the long-lived NLSP.
$\tilde{e}_{R}$ mainly decays into a SM electron and a gravitino (main mode).
However, due to its smuon component, it also decays into a SM muon and a gravitino (rare mode).
Hence we expect to observe rare mode events
 where one of the sparticle pair produced by the collider decays into a SM muon
 and the other into a SM electron
 with large vertex separation due to the longevity of $\tilde{e}_{R}$.

The bulk matter RS model predicts that 
 the branching ratio of the rare mode is given by
\begin{eqnarray}
Br(\tilde{e}_{R} \ \rightarrow \ \mu \ \psi_{3/2})
&\sim& \left( \frac{ \epsilon_{1} }{ \epsilon_{2} } \right)^{2}
 \ + \ \left( \epsilon_{1} \delta_{2} \frac{v_{d}}{M_{susy}} \right)^{2} \nonumber
\\
&\sim& \left( 3 \frac{ m_{e} }{ m_{\mu} } \right)^{2}
 \ + \ \left( 3 \frac{ m_{e} }{M_{susy}} \right)^{2} \nonumber
\\
&\sim& 0.0002 \ ,
\end{eqnarray}
 where we neglected the second terms of the right hand sides
 because we have 
\begin{eqnarray*}
3 \frac{ m_{e} }{ m_{\mu} } \ >> \ 3 \frac{ m_{e} }{M_{susy}}
\end{eqnarray*}
 in realistic SUSY models.

On the other hand, the MFV scenario predicts that
 the branching ratio of the rare mode is given by
\begin{eqnarray}
Br(\tilde{e}_{R} \ \rightarrow \ \mu \ \psi_{3/2}) 
 &\sim& \left\{ 6 (\zeta_{3})^{2} (\delta_{1})^{2} \frac{ \epsilon_{1} }{ \epsilon_{2} } \right\}^{2}
 \ + \ \left\{ 2 (\zeta_{3})^{2} (\delta_{1})^{2} \epsilon_{2} \delta_{1} \frac{v_{d}}{M_{susy}} \right\}^{2} \nonumber
\\
&\sim& (\zeta_{3} \delta_{1})^{4} \times 0.03 \ ,
\end{eqnarray}
Again, the bound on $\mu \rightarrow e \gamma$ branching ratio gives a severe constraint 
 on the value of $\zeta_{3} \delta_{1}$,
 and the branching ratio satisfies
\begin{eqnarray}
Br(\tilde{e}_{R} \ \rightarrow \ \mu \ \psi_{3/2}) 
&<& 10^{-6}
\end{eqnarray}
 for realistic mass spectra.

$\tilde{e}_{R}$ also decays into a SM tau and a gravitino 
 but the branching ratio is suppressed by the factor $(m_{e}/m_{\tau})^{2}$
 and is thus negligibly small.
\\

Next consider the case where $\tilde{\mu}_{R}$ is lighter than $\tilde{e}_{R}$
 and is the long-lived NLSP.
$\tilde{\mu}_{R}$ mainly decays into a SM muon and a gravitino (main mode),
 but also into a SM electron and a gravitino,
 or into a SM tau and a gravitino (rare modes).
The branching ratio of the rare mode where the sparticle pair 
 decay into a muon and an electron and two gravitinos is the same
 as (92, 93).
The branching ratio of the rare mode where the sparticle pair 
 decay into a tau and a muon and two gravitinos is the same
 as (89, 90).
\\

\subsection{Type III - Scharm Rare Decay into SM Top}

\subsubsection{Scharm is Lighter than Stop}

\ \ \ Consider the case where $\tilde{c}_{R}$ is lighter than $\tilde{t}_{1}$.
Tune the center-of-mass energy between the thresholds of 
 $\tilde{c}_{R}$ and $\tilde{t}_{1}$.
Then $\tilde{c}_{R}, \tilde{u}_{R}, \tilde{s}_{R}, \tilde{d}_{R}$,
 whose masses are almost degenerate,
 are produced on-shell,
 while other squark mass eigenstates are not.
$\tilde{c}_{R}$ / $\tilde{u}_{R}$ / $\tilde{s}_{R}$ / $\tilde{d}_{R}$
 mainly decay into SM charm / up / strange / down and the lightest neutralino
 $\chi^{0}_{1}$, which is Bino-like (main mode).
$\chi^{0}_{1}$ promptly decays into several SM leptons and NLSP,
 e.g. we have
\begin{eqnarray}
e e \ \rightarrow \ \tilde{c}_{R} \ \tilde{c}_{R}
\ \rightarrow \ c \ \chi^{0}_{1} \ c \ \chi^{0}_{1}
\ \rightarrow \ (c{\rm -jet}) \ NLSP \ (c{\rm -jet}) \ NLSP \ ({\rm SM \ leptons})
\end{eqnarray}
 for scharm pair production.
Due to the small stop components, 
 they also decay into SM top and $\chi^{0}_{1}$
 with a tiny branching ratio (rare mode), e.g. we have
\begin{eqnarray}
e e \ \rightarrow \ \tilde{c}_{R} \ \tilde{c}_{R}
\ \rightarrow \ t \ \chi^{0}_{1} \ c \ \chi^{0}_{1}
\ \rightarrow \ ({\rm top \ decay \ products}) \ NLSP \ (c{\rm -jet}) \ NLSP \ ({\rm SM \ leptons})
\end{eqnarray}
 for scharm pair production.
The signal of the main mode is two hard jets,
 two long-lived charged massive particles and several SM leptons.
On the other hand, the signal of the rare mode is, when SM top decays hadronically,
  four hard jets, two long-lived charged massive particles and several SM leptons.
We see that the probability of misidentifying a main mode event
 as a rare mode event is negligibly small
 if we require the hadronic top decay in rare mode events.

Of the four eigenstates,
 $\tilde{c}_{R}$ dominantly contributes to rare mode events
 because the stop components in the other eigenstates are more suppressed 
 than in $\tilde{c}_{R}$.
On can confirm this by
 requiring $c$-tagging for one of the jets in rare mode events.

From (80) with ($Q, \alpha$) replaced by ($U, \beta$) and (81),
 the bulk matter RS model predicts that the branching ratio of the rare mode
 is given by
\begin{eqnarray}
Br(\tilde{c}_{R} \ \rightarrow \ t \ \chi^{0}_{1}) 
 &\sim& \left( \frac{ \beta_{2} }{ \beta_{3} } \right)^{2} 
 \ + \ \left( \beta_{2} \alpha_{3} \frac{ v_{u} }{ M_{susy} } \right)^{2} \nonumber
\\
 &\sim& \left( \frac{1}{\lambda^{2}} \frac{m_{c}}{m_{t}} \right)^{2} 
 \ + \ \left( \frac{1}{\lambda^{2}} \frac{m_{c}}{m_{t}} 
 \ \frac{ m_{t} }{ M_{susy} } \right)^{2} \nonumber
\\
&\sim& 0.02 \ ,
\end{eqnarray}
 where we neglected the second term because we have
\begin{eqnarray*}
m_{t} < M_{susy}
\end{eqnarray*}
 in realistic SUSY models.

From (84) and (85),
 the MFV scenario predicts that 
 the branching ratio of the rare mode is given by
\begin{eqnarray}
Br(\tilde{c}_{R} \ \rightarrow \ t \ \chi^{0}_{1}) 
 &\sim& \left\{ 6 (\gamma_{3})^{2} (\alpha_{2})^{2} \frac{ \beta_{2} }{ \beta_{3} } \right\}^{2} 
 \ + \ \left\{ 2 (\gamma_{3})^{2} (\alpha_{2})^{2} \beta_{2} \alpha_{3} \frac{ v_{u} }{ M_{susy} } \right\}^{2} \nonumber
\\
 &\lesssim& 5 \times 10^{-6} \ ,
\end{eqnarray}
 where we used the fact that $\gamma_{3} \leq 1$.
Comparing (96) with (95),
 we notice that the bulk matter RS model and the MFV scenario
 have distinctively different predictions.
\\

\subsubsection{Stop is Lighter}

\ \ \ Consider the case where $\tilde{t}_{1}$ is lighter than $\tilde{c}_{R}$.
$\tilde{c}_{R}$ is still lighter than ``almost SU(2) doublet squark" mass eigenstates (including $\tilde{t}_{1}, \tilde{b}_{1}$).
Tune the center-of-mass energy between 
 the thresholds of $\tilde{c}_{R}$ and ``almost doublet squark" eigenstates.
Then $\tilde{c}_{R}, \tilde{u}_{R}, \tilde{s}_{R}, \tilde{d}_{R}, 
 \tilde{b}_{1}, \tilde{t}_{1}$ are produced on-shell,
 while the other squark mass eigenstates are not.

We want to extract the signals of rare mode events where one of the pair of
 $\tilde{c}_{R}, \tilde{u}_{R}, \tilde{s}_{R}$ or $\tilde{d}_{R}$s
 decays into a SM top and a neutralino.
However these events are contaminated by
 the events where one of the pair of $\tilde{b}_{1}$s decays into
 a SM top and a chargino;
 the chargino decays into a NLSP and SM leptons,
 but one charged lepton is mis-detected.
$b$-jet from the other $\tilde{b}_{1}$ is mis-$b$-tagged.
There is also a contamination from the events where one of the pair of $\tilde{t}_{1}$s
 decays into a SM bottom and a chargino, 
 or into a SM charm and a neutralino due to the scharm component in $\tilde{t}_{1}$.
We take advantage of kinematical properties to reject these contaminations.

Tune the center-of-mass energy close to the threshold of 
 $\tilde{c}_{R}, \tilde{u}_{R}, \tilde{s}_{R}, \tilde{d}_{R}$
 so that they are produced almost at rest.
Suppose that one observed three hard jets from the hadronic decay of SM top ($t$), 
 another hard jet ($j$)
 and several SM leptons plus two NLSPs.
Further assume that the 3-momenta of $t$ and $j$ are reconstructed successfully.
We want to know whether this event comes from the decay of 
 $\tilde{c}_{R}, \tilde{u}_{R}, \tilde{s}_{R}, \tilde{d}_{R}$
 or from $\tilde{t}_{1}, \tilde{b}_{1}$.
In the former case,
 the 3-momenta satisfy
\begin{eqnarray}
\vert \vec{p}_{j} \vert \ + \ \sqrt{ \vert -\vec{p}_{j} \vert^{2} + m_{\chi}^{2} }
 &\simeq& m_{\tilde{c}_{R}} \ ,
\\
\sqrt{ \vert \vec{p}_{t} \vert^{2} + m_{t}^{2} } \ + \ \sqrt{ \vert -\vec{p}_{t} \vert^{2} + m_{\chi}^{2} }
 &\simeq& m_{\tilde{c}_{R}} \ ,
\end{eqnarray}
 where $\vec{p}_{j}$ and $\vec{p}_{t}$ respectively denote the 3-momenta of $j$ and $t$,
 and $m_{\chi}$ the mass of Bino-like neutralino.
In the latter case, however,
 the above equations hold for specific situations
 where $j$ and $t$ go in special directions against the initial $\tilde{t}_{1}$s or $\tilde{b}_{1}$s
 because the $\tilde{t}_{1}, \tilde{b}_{1}$s are boosted.
To summarize, the rare mode signals of $\tilde{c}_{R}, \tilde{u}_{R}, \tilde{s}_{R}, \tilde{d}_{R}$
 can be extracted through the discriminants (99, 100).

The branching ratios of the rare mode in the bulk matter RS model 
 and the MFV scenario are the same as in (97, 98).

\subsection{Cross Sections}

\ \ \ We calculate the cross sections of the rare modes given above
 and discuss their accessibility at collider experiments.

First focus on the Type I experiment, where one of the pair of ``almost singlet smuon or selectron mass eigenstates" 
 decays into two SM taus and a NLSP
 in case the NLSP is stau-like.
The center-of-mass energy is tuned above the threshold of $\tilde{\mu}_{R}$, namely
 we take
\begin{eqnarray*}
\sqrt{s} &=& 2 m_{\tilde{\mu}_{R}} \ + \ 100 \ {\rm GeV} \ .
\end{eqnarray*}
In Figure 1, we plot the mass of $\tilde{\mu}_{R}$ vs. the cross section of the rare mode
 at a $e^{+}e^{-}$ collider.
We take the branching ratio of the rare mode as
\begin{eqnarray*}
Br(\tilde{\mu}_{R} \ \rightarrow \ \tau \ \tau \ \tilde{\tau}_{1}) &=& \left( \frac{m_{\mu}}{m_{\tau}} \right)^{2}
\end{eqnarray*}
 based on (88).
We assume that $\tilde{\mu}_{R}$ is slightly lighter than $\tilde{e}_{R}$
 so that a half of $\tilde{e}_{R}$s decay into $\tilde{\mu}_{R}$
 and contribute to the rare mode.
In calculating the $\tilde{e}_{R}$ pair production cross section,
  the Bino mass is assumed to be $1.5 m_{\tilde{\mu}_{R}}$.
Also shown is the total cross section of
 $\tilde{\mu}_{R}, \tilde{e}_{R}$ production events.

\begin{figure}[htbp]
 \begin{center}
  \includegraphics[width=100mm]{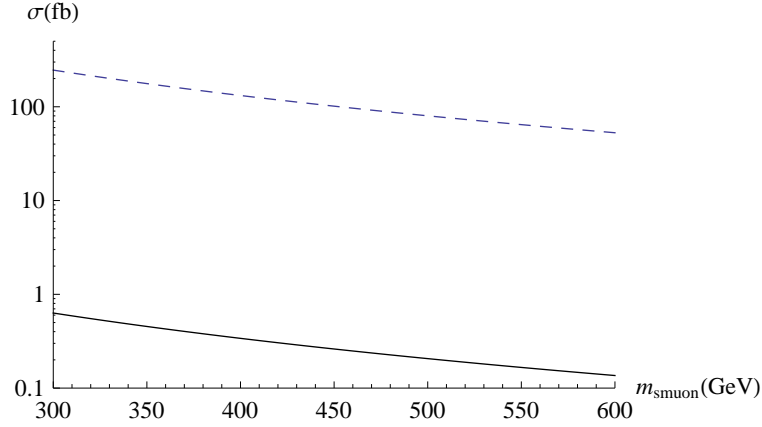}
 \end{center}
 \caption{
 Mass of $\tilde{\mu}_{R}$ vs. the cross section of the rare mode
  where a $\tilde{\mu}_{R}$ is produced and decays into two SM taus and a NLSP stau
  (the straight line).
 The center-of-mass energy is taken as $\sqrt{s} = 2 m_{\tilde{\mu}_{R}} + 100$ GeV.
 The total cross section of $\tilde{\mu}_{R}$ or $\tilde{e}_{R}$ production process
  is also shown (the dashed line).
 }
\end{figure}

Next focus on the Type II, where one of the pair of ``almost singlet selectron mass eigenstates" 
 decays into a SM muon and a gravitino if it is the NLSP,
 or one of the pair of ``almost singlet smuon mass eigenstates" 
 decays into a SM electron / tau and a gravitino if it is the NLSP.
We tune the center-of-mass energy slightly
 above the threshold of $\tilde{e}_{R}$ / $\tilde{\mu}_{R}$
 so that their velocities are low enough to trap them inside the inner detector.
For simplicity, we take
\begin{eqnarray*}
\sqrt{s} &=& 2 m_{\tilde{e}_{R} / \tilde{\mu}_{R}} \ + \ 20 \ {\rm GeV} \ ,
\end{eqnarray*}
In Figure 2, we plot the mass of $\tilde{e}_{R}$ vs. the cross section of the rare mode
 if it is the NLSP,
and the mass of $\tilde{\mu}_{R}$ vs. the cross section of the rare mode
 if it is the NLSP.
We take the branching ratios of the rare modes as
\begin{eqnarray*}
Br(\tilde{e}_{R} \ \rightarrow \ \mu \ \psi_{3/2}) &=&
 Br(\tilde{\mu}_{R} \ \rightarrow \ e \ \psi_{3/2}) \ = \
 \left( \frac{1}{3} \frac{m_{e}}{m_{\mu}} \right)^{2} \ ,
\\
Br(\tilde{\mu}_{R} \ \rightarrow \ \tau \ \psi_{3/2}) &=&
 \left( \frac{m_{\mu}}{m_{\tau}} \right)^{2} \ ,
\end{eqnarray*}
 based on (88, 92).
Also shown are the total cross sections of 
 $\tilde{e}_{R}$ and $\tilde{\mu}_{R}$ production processes.

\begin{figure}[htbp]
 \begin{minipage}{0.5\hsize}
  \begin{center}
   \includegraphics[width=80mm]{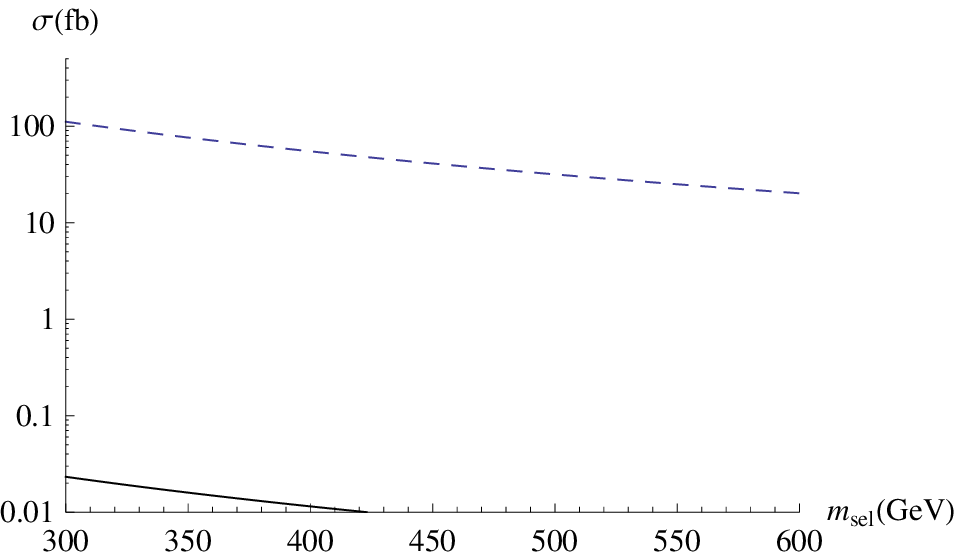}
  \end{center}
 \end{minipage}
 \begin{minipage}{0.5\hsize}
  \begin{center}
   \includegraphics[width=80mm]{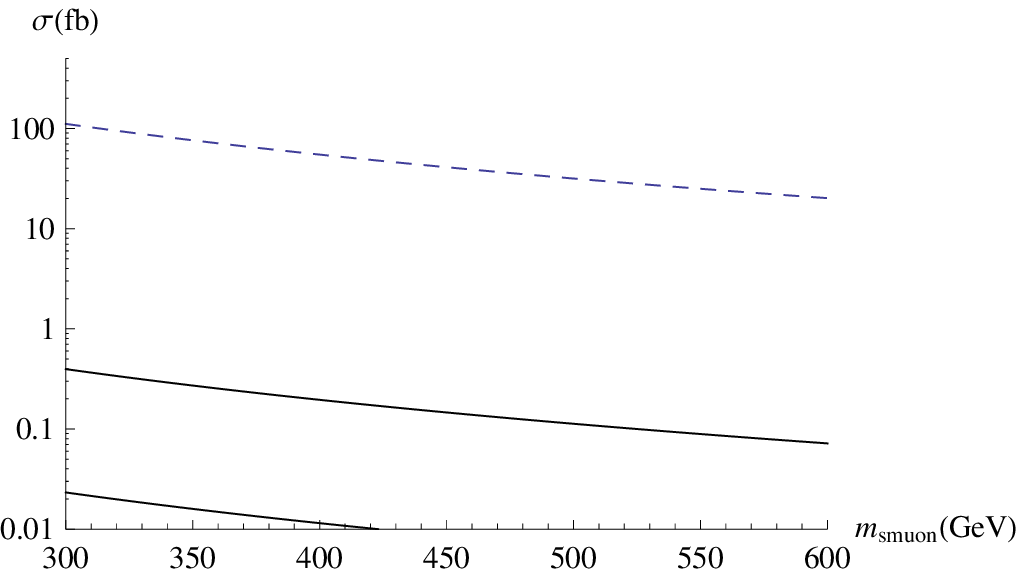}
  \end{center}
 \end{minipage}
 \caption{
 Left: Mass of $\tilde{e}_{R}$ vs. the cross section of the rare mode
  where a NLSP $\tilde{e}_{R}$ is produced and decays into a SM muon and a gravitino
  (the straight line).
 Right: Mass of $\tilde{\mu}_{R}$ vs. the cross section of a rare mode
  where a NLSP $\tilde{\mu}_{R}$ is produced and decays into a SM electron and a gravitino
  (the straight line below), and the other rare mode
  where a NLSP $\tilde{\mu}_{R}$ is produced and decays into a SM tau and a gravitino
  (the straight line above).
 The total cross sections of $\tilde{e}_{R}$ (left) and $\tilde{\mu}_{R}$ (right)
  production processes are also shown (the dashed lines).
  }
\end{figure}

Finally focus on the Type III, where one of the pair of 
 $\tilde{c}_{R}, \tilde{u}_{R}, \tilde{s}_{R}, \tilde{d}_{R}$s
 decays into a SM top and a neutralino.
The center-of-mass energy is tuned above the threshold of $\tilde{c}_{R}$.
First we take
\begin{eqnarray*}
\sqrt{s} &=& 2 m_{\tilde{c}_{R}} \ + \ 100 \ {\rm GeV} \ 
\end{eqnarray*}
 so that the cross section is nearly maximized.
Second we take
\begin{eqnarray*}
\sqrt{s} &=& 2 m_{\tilde{c}_{R}} \ + \ 10 \ {\rm GeV} \ 
\end{eqnarray*}
 so that $\tilde{c}_{R}$s
 are produced almost at rest and 
 the rare mode events are kinematically distinguishable from
 $\tilde{t}_{1}, \tilde{b}_{1}$ production events in case $\tilde{t}_{1}, \tilde{b}_{1}$ are lighter.
In Figure 3, we plot the mass of $\tilde{c}_{R}$ vs. the cross section of the rare mode
 for both cases.
The branching ratio of the rare mode is taken as
\begin{eqnarray*}
Br(\tilde{c}_{R} \ \rightarrow \ t \ \chi_{1}^{0}) &=&
 \left( \frac{1}{\lambda^{2}} \frac{m_{c}}{m_{t}} \right)^{2}
\end{eqnarray*}
 for both cases, based on (97).
Also shown is the total cross section of 
 $\tilde{c}_{R}, \tilde{u}_{R}, \tilde{s}_{R}, \tilde{d}_{R}$ production processes.

\begin{figure}[htbp]
 \begin{minipage}{0.5\hsize}
  \begin{center}
   \includegraphics[width=90mm]{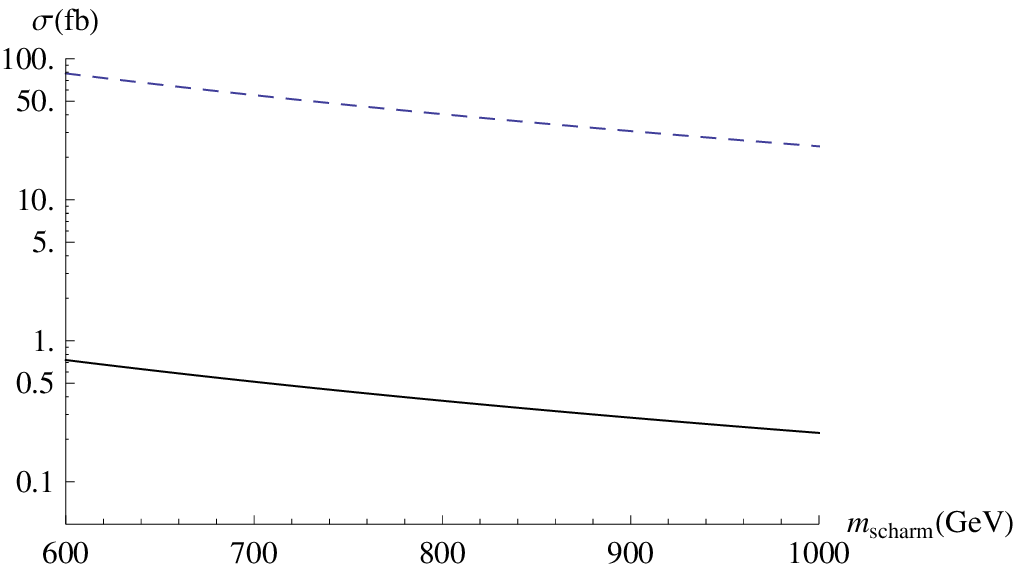}
  \end{center}
 \end{minipage}
 \begin{minipage}{0.5\hsize}
  \begin{center}
   \includegraphics[width=90mm]{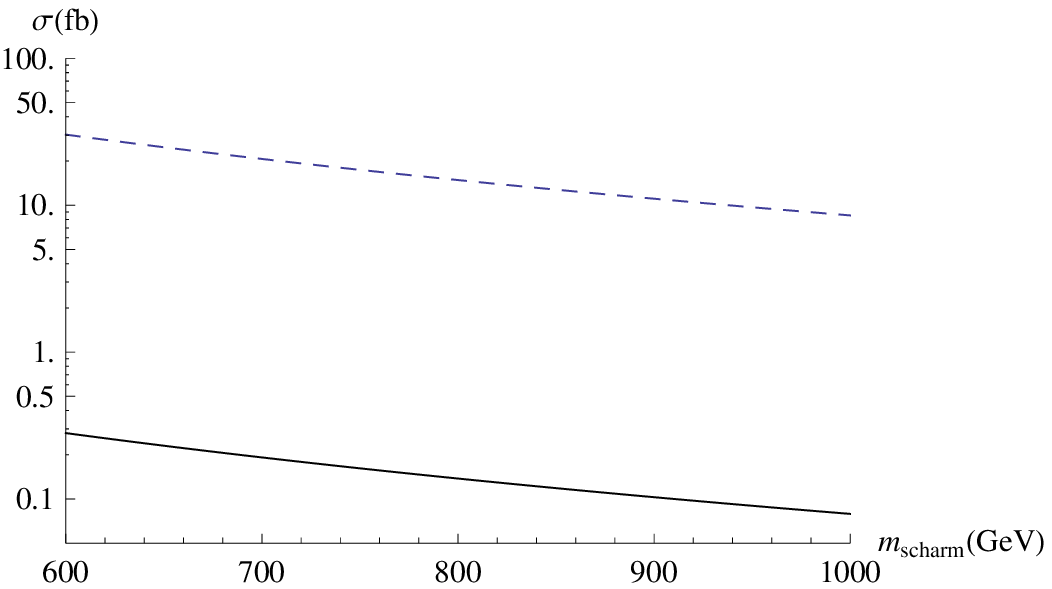}
  \end{center}
 \end{minipage}
 \caption{
 Mass of $\tilde{c}_{R}$ vs. the cross section of the rare mode
  where a $\tilde{c}_{R}$ is produced and decays into a SM top and a neutralino
  (the straight lines).
 The center-of-mass energy is taken as $\sqrt{s} = 2 m_{\tilde{c}_{R}} + 100$ GeV
  in the left and $\sqrt{s} = 2 m_{\tilde{c}_{R}} + 10$ GeV
  in the right.
 The total cross section of $\tilde{c}_{R}, \tilde{u}_{R}, \tilde{s}_{R}$ or $\tilde{d}_{R}$ 
  production processes is also shown (the dashed lines).
  }
\end{figure}

Note that we can in principle reject all background events
 when observing the signals of the rare modes.
Therefore detecting several signals is sufficient to confirm the bulk matter RS model.
From Figure 1 and 2, we find that one can study
 the stau component in $\tilde{\mu_{R}}$
 at the ILC with the integrated luminosity of $\sim 100$fb$^{-1}$.
However, studying the smuon component in $\tilde{e_{R}}$
 or the selectron component in $\tilde{\mu_{R}}$
 requires $\sim 1000$fb$^{-1}$ integrated luminosity.
From Figure 3, we see that
 $\sim 100$fb$^{-1}$ integrated luminosity is sufficient to study
 the stop component in $\tilde{c_{R}}$.
\\

\section{Summary and Outlook}

\ \ \ We discussed observing signals of the bulk matter RS model,
 especially when the IR scale is far above TeV scale.
We saw that this is possible in the case of 
 the minimal supersymmteric extension of the bulk matter RS model
 where the warped spacetime solely explains the hierarchy of the Yukawa couplings,
 while SUSY solves the gauge hierarchy problem.
There, gravity mediation contributions to soft SUSY breaking terms 
 reflect the 5D disposition of superfields.
Hence flavor-violating soft SUSY breaking matter mass terms that arise from gravity mediation
 exhibit a flavor structure unique to the bulk matter RS model.
RG running of the Yukawa coupling constants also contributes to the flavor-violating terms,
 but its contributions and the gravity mediation contributions are distinguishable
 if the mass scale of gauge mediation is not much larger than that of gravity mediation.
Then the latter contributions can be extracted
 by investigating the 1st and 2nd generation SU(2) singlet SUSY particles,
 where the former are further suppressed by the small Yukawa coupling constants.
We focused on the flavor compositions of SUSY particle mass eigenstates,
 which reflect the relative size of flavor-violating soft SUSY breaking terms.
We enumerated three modes of collider experiments
 where one can measure the compositions 
 by observing rare decays of SUSY particles.
Predictions on their branching ratios were made based on the bulk matter RS model,
 and were compared with those of the minimal flavor violation scenario.
These predictions will be confirmed or rejected by a future lepton collider 
 whose center-of-mass energy is tuned appropriately.

A lesson of this study is that
 if new physics at TeV scale contains a flavor-violating sector other than the Yukawa couplings,
 it is possible to observe signatures of models that explain the Yukawa coupling hierarchy
 through the flavor structure of the new sector.
In the case of this paper, MSSM contains gravity-mediation-origined soft mass terms,
 which provide a new source of flavor violation.
Gravity mediation and the Yukawa couplings are independent in the original MSSM,
 but have a correlation if the bulk matter RS model is the origin of the Yukawa coupling hierarchy.
Hence we can predict the flavor structure of gravity mediation contributions
 (up to their orders of magnitudes) from the data on SM,
 and eventually confirm or reject the bulk matter RS model 
 through a detailed study on SUSY matter particles.
This study can be extended to any new physics scenario at TeV scale
 as long as it couples to matter fields and may violate flavor.
In any case, SU(2) singlet muon and charm and their new physics partners
 play a pivotal role;
SU(2) singlets receive less flavor-violating quantum corrections from the SM Yukawa couplings,
 and thus new flavor-violating terms are easy to extract.
Since the 1st and 2nd generation particles only have small Yukawa couplings,
 we expect that their new SU(2) singlet partners almost do not mix with SU(2) doublets.
Muon has much larger Yukawa coupling than electron and is much more sensitive to
 the origin of the Yukawa coupling hierarchy.
The partner of charm may have a large flavor-violating mixing with top,
 the only quark whose flavor can be identified with virtually no misidentification rate.
\\

\section{Acknowledgements}

\ \ \ TY is grateful to Yasuhiro Okada (KEK), Nobuchika Okada (The University of Alabama)
 and Keisuke Fujii (KEK) for useful advice on writing this paper.
This work is in part supported by a grant of the Japan Society for the Promotion of Science,
 No. 23-3599.
\\

\end{document}